\documentstyle[aps,epsfig,multicol,fancyheadings,amssymb]{revtex}
\begin{document}
%%%%%%%%%%%%%%%%%%%%%%%%%%%%%%%%%%%%%%%%%%%%%%%%%%%%%%%%

\title{Susceptibility and Percolation in 2D Random Field Ising Magnets}

\author{E.\ T.\ Sepp\"al\"a and M.\ J.\ Alava}

\address{Helsinki University of Technology, Laboratory of Physics, 
P.O.Box 1100, FIN-02015 HUT, Finland}
\date{\today}

\maketitle

\begin{abstract}
The ground state structure of the two-dimensional random field Ising
magnet is studied using exact numerical calculations. First we show
that the ferromagnetism, which exists for small system sizes, vanishes
with a large excitation at a random field strength dependent length
scale. This {\it break-up length scale} $L_b$ scales exponentially
with the squared random field, $\exp(A/\Delta^2)$. By adding an
external field $H$ we then study the susceptibility in the ground
state.  If $L>L_b$, domains melt continuously and the magnetization
has a smooth behavior, independent of system size, and the
susceptibility decays as $L^{-2}$. We define a random field strength
dependent critical external field value $\pm H_c(\Delta)$, for the up
and down spins to form a percolation type of spanning cluster. The
percolation transition is in the standard short-range correlated
percolation universality class. The mass of the spanning cluster
increases with decreasing $\Delta$ and the critical external field
approaches zero for vanishing random field strength, implying the
critical field scaling (for Gaussian disorder) $H_c \sim (\Delta
-\Delta_c)^\delta$, where $\Delta_c = 1.65 \pm 0.05$ and
$\delta=2.05\pm 0.10$. Below $\Delta_c$ the systems should percolate
even when $H=0$.  This implies that even for $H=0$ above $L_b$ the
domains can be fractal at low random fields, such that the largest
domain spans the system at low random field strength values and its
mass has the fractal dimension of standard percolation $D_f = 91/48$.
The structure of the spanning clusters is studied by defining {\it red
clusters}, in analogy to the ``red sites'' of ordinary
site-percolation. The size of red clusters defines an extra length
scale, independent of $L$.
\end{abstract}

\noindent {\it PACS \# \ \ 05.50.+q, 75.60.Ch, 75.50.Lk, 64.60.Ak}

\begin{multicols}{2}[]
\narrowtext

\section{Introduction}\label{intro}

The question of the importance of quenched random field (RF) disorder
in ferromagnets traces back to the primary paper by Imry and
Ma~\cite{Imr75,Young97} from mid-seventies. They argued using energy
minimization for an excitation to the ground state, that the
randomness in the fields assigned to spins changes the lower critical
dimension from the pure case with $d_l=1$ to $d_l=2$. After that a
number of field-theoretical calculations suggested that the randomness
increases $d_l$ with two to be $d_l=3$.  Finally came rigorous proofs
first by Bricmont and Kupiainen~\cite{Bri87} in '87 that there is a
ferromagnetic phase in the three-dimensional (3D) random field Ising
model (RFIM) and in '89 by Aizenman and Wehr~\cite{Aiz89} that there
is no ferromagnetic phase in 2D RFIM. Thus it was established that the
lower critical dimension is two. This means that the ground state is a
paramagnet, but the problem as to how to describe the structure of the
(ground state of) 2D RFIM still persists. Some recent work concerns
the scaling of the correlation lengths~\cite{stinch} and there is a
suggestion of a ferromagnetic phase, but with a magnetization that is
in the thermodynamic limit below unity~\cite{Vives99}. The point is
that due to the (relevant) disorder there are no easy arguments that
would indicate, say, how the paramagnetic ground state should be
characterized. This is different from the thermal Ising case, which is
quite trivial in 1D.

In two dimensional Ising magnets, in the presence of quenched random
fields, the problem of determining the ground state (GS) becomes more
difficult. Finding the true ground state with any standard Monte Carlo
method is problematic due to the complex energy landscape. Even with
the exact ground state methods, as the one used in this paper, the
thermodynamic limit is difficult to reach, since the finite size
effects are strong. In the typical case of square lattices the only
way not to have a massive domain, which would scale with the Euclidean
dimension of the system, is to have enough interpenetrating domains of
both spin orientations.  However, with decreasing strength of the
randomness the ferromagnetic coupling constants between spins start to
matter, the domains become ``thicker'', and thus one enters an
apparent ferromagnetic regime, and the paramagnetic (PM) phase is
encountered only at very large length scales. Should there be large
clusters with a fractal (non-Euclidian) mass scaling they nevertheless
can contribute to the physics in spite of the fact that the total
fraction of spins can be negligible in the thermodynamics limit. Thus
such clusters may even be measurable in experiments or be related to
the dynamical behavior in non-equilibrium conditions. Therefore it is
of interested to study the structure of the large(st) clusters in the
ground state, since it is not simply paramagnetic like in normal Ising
magnets above $T_c$. The true ground state structure gives also some
insight into the physics at $T>0$, since the overlap between the GS
and the corresponding finite-$T$ state is close to unity for $T$
small, in contrast to the thermal chaos in spin glasses~\cite{chaos}.

In this paper we want to shed some light on the character of the
ground states of 2D RFIM. We have done extensive exact ground state
calculations in order to characterize how the ferromagnetic (FM) order
vanishes with increasing system size.  We have also studied the effect
of the application of an external field, that is the susceptibility of
the 2D RFIM. Allowing for a non-zero external field makes it possible
to investigate a percolation -type of critical phenomenon for the
largest clusters. We propose a phase diagram in the disorder strength
and external field plane for the percolation behavior. The presence of
clusters of the size of the system, i.e. percolation type of order
brings another correlation length in the systems and thus makes the
decay of ferromagnetic order more complicated than at first sight.

The Hamiltonian of the random field Ising model is
\begin{equation}
{\mathcal H} = -J \sum_{\langle i j \rangle}S_i S_j - \sum_i 
(h_i + H) S_i,
\label{eq:H}
\end{equation}
where $J>0$ (in this paper we use $J=1$ for numerical calculations) is
the coupling constant between nearest-neighbor spins $S_i$ and
$S_j$. We use here square-lattice. $H$ is a constant external field,
which if non-zero is assigned to all of the spins, and $h_i$ is the
random field, acting on each spin $S_i$. We consider mainly a
Gaussian distribution for the random field values
\begin{equation}
P(h_i) = \frac{1}{\sqrt{2 \pi} \Delta} \exp\left[-\frac{1}{2}
\left(\frac{h_i}{\Delta}\right)^2\right], 
\label{eq:Gaussian}
\end{equation}
with the disorder strength given by $\Delta$, the standard deviation 
of the distribution, though in some cases the bimodal distribution,
\begin{equation}
P(h_i) = \frac{1}{2} \delta (h_i -\Delta) + \frac{1}{2} 
\delta (h_i +\Delta)
\label{eq:bimodal}
\end{equation}
is used, too. The results presented below should not be too
dependent on the actual $P(h)$, in any case.

To find the ground state structure of the RFIM means that the
Hamiltonian (\ref{eq:H}) is minimized, in which case the positive
ferromagnetic coupling constants prefer to have all the spins aligned
to the same direction. On the other hand the random field contribution
is to have the spins to be parallel with the local field, and thus has
a paramagnetic effect. This competition of ferromagnetic and
paramagnetic effects leads to a complicated energy landscape and the
finding the GS becomes a global optimization problem.  An interesting
side of the RFIM is that it has an experimental realization as a
diluted antiferromagnet in a field (DAFF). By gauge-transforming the
Hamiltonian of DAFF
\begin{equation}
{\mathcal H} = -J \sum_{\langle i j \rangle}S_i S_j \epsilon_i \epsilon_j 
- B \sum_i \epsilon_i S_i,
\label{eq:DAFF}
\end{equation}
where the coupling constants $J<0$, $\epsilon_i$ is the occupation
probability of a spin $S_i$, and $B$ is now a constant external field,
one gets the Hamiltonian of RFIM~(\ref{eq:H}) with $H=0$
\cite{Fish79,Cardy,Belanger97}. The ferromagnetic order in
the RFIM corresponds to antiferromagnetic order in the DAFF,
naturally.

As background, it is of interest to review a few basic results.  Imry
and Ma used a domain-wall argument to show that the lower critical
dimension $d_l=2$~\cite{Imr75}. In order to have a domain there is an
energy cost of ${\mathcal O}(L^{d-1})$ from the domain wall. On the
other hand the system gains energy by flipping the domain from the
fluctuations of random fields, which interpreted as a typical
fluctuation means that the gain is ${\mathcal O}(L^{d/2})$.  Thus,
whenever $d/2 \geq d-1$, i.e. $d \leq 2$, it is energetically
favorable for the system to break into domains.  However, in this
paper we will point out, as it has been shown in 1D~\cite{gerd}, that
the ${\mathcal O}(L^{d/2})$ scaling can be used only in relation to
the sum of the random fields in the ``first excitation'', but not to the
droplet field energy when the GS consists of domains of different
length scales.

Grinstein and Ma~\cite{GMa} derived from the continuum interface
Hamiltonian that the roughness of the domain wall (DW) in RFIM scales as $w
\sim \Delta^{2/3} L^{(5-d)/3}$, which is consistent with
$d_l=2$.  Later Fisher~\cite{Fis86} used the functional
renormalization group (FRG) to obtain the roughness exponent $\zeta
=(5-d)/3$ and argued that due to the existence of many metastable
states the perturbative RG calculations and dimensional reduction
fail. Another, microscopic calculation by Binder~\cite{Bin83}
optimized the domain-wall energy in two-dimensions.  The net result is
a total energy gain from random fields, $\Delta U = - (\Delta^2/J) L
\ln L$ due to domain wall decorations, which implies that the domain
wall energy $U = 2JL + \Delta U$ vanishes on a minimal length scale
\begin{equation}
L_b \sim \exp[A(J/\Delta)^2],
\label{eq:L_b}
\end{equation}
where $A$ is a constant of order unity.  For $L > L_b$ the expectation
is that the system spontaneously breaks up into domains.  Similarly
the energy of a domain with a constant external field $H$ becomes
\begin{equation}
U/J = 8L + 2 (H/J)L^2 - (8/A) (\Delta/J)^2L \ln L.
\label{eq:U_H}
\end{equation}
Setting $U/J =0$ and assuming that the critical length scale $L_{b,H}$
scales as $L_b$, without the field, i.e. $L_{b,H} \sim L_b$, the
critical external field becomes
\begin{equation}
H_c/J = (4/A)(\Delta/J)^2 \exp[-A(J/\Delta)^2 -1].
\label{eq:Hc_binder}
\end{equation}
Note, that in this case the first two terms in (\ref{eq:U_H}) 
assume that the domain is compact.

These results imply, together with the notion that the ground state is
paramagnetic, that the magnetization should not as such display any
``universal'' features.  The results of this paper show that the
magnetization is {\em not} dependent on the system size and has a
smooth behavior of $m \sim f[H/\exp(-6.5/\Delta)]$ and the
susceptibility vanishes with the system size as $\chi \sim L^{-2}
\exp(7.3/\Delta) g[H/\exp(-6.5/\Delta)]$, where $g(y \simeq 0) \simeq$
const, and $g(y \to \pm \infty) \sim \exp(-0.2|y|)$. These imply that
there is a length scale, related to the rate at which clusters
``melt'' when $H$ is changed from zero.

The presence of such a length scale is qualitatively similar to the one
discovered in the context of the percolation transition.  It turns out
so that when the external field is varied, the universality class is
that of the ordinary short-range correlated percolation universality
class. The external field threshold for spanning $H_c$ with respect to
decreasing random field strength approaches zero external field limit
from the site-percolation limit of infinite random field strength
value, suggesting a behavior for Gaussian disorder of $H_c \sim
(\Delta -\Delta_c)^\delta$, where $\Delta_c = 1.65 \pm 0.05$ and
$\delta=2.05\pm 0.10$. Below this value the lattice effects of site
percolation are washed out, and there is yet another length scale that
characterizes the percolation clusters, the size of the ``red
clusters'' defined below in analogy to the usual red or cutting sites
in percolation. Now a whole cluster is reversed due to the forced
reversal of a ``seed'' cluster, when the sample is optimized
again. The length scale is however finite, indicating that the global
optimization of the ground state creates only finite spin-spin
correlations as is the case in 1D as well.

This paper is organized so that it starts by introducing in
Section~\ref{method} the exact ground state calculation technique. In
Section~\ref{breakup} the breaking up of ferromagnetic order is
discussed, based on a nucleation of droplets -picture which follows
from a level crossing between a FM ground state and one with a large
droplet. The relevant $L_b$ scaling (\ref{eq:L_b}) is derived from
extreme statistics.  Above the break-up length scale the domains have
a complex structure that is briefly discussed.  The effect of an
external field, in the case the system size is above $L_b$, is studied
in Section~\ref{suskis} for Gaussian disorder.  The percolation
aspects of the 2D RFIM are studied in detail in
Section~\ref{perkolaatio_H}.  The phase diagram for the percolation
behavior as functions of the external field and the random field
strength is sketched and the properties of the transition are
discussed. The zero-external field percolation probability is studied
in Section~\ref{perkolaatio_0}. In the section also the structure of
spanning clusters is studied using the so called {\em red clusters}
whose scaling and properties are discussed.  The paper is finished
with conclusions in Section~\ref{concl}.

\section{Numerical Method}\label{method}

For the numerical calculations the Hamiltonian (\ref{eq:H}) is
transformed to a random flow graph with two extra sites: the source
and the sink. The positive field values $h_i$ correspond flow
capacities $c_{it}$ connected to the sink ($t$) from a spin $S_i$,
similarly the negative fields with $c_{is}$ are connected to the
source ($s$), and the coupling constants $J_{ij} \equiv 2c_{ij}$
between the spins correspond flow capacities $c_{ij} \equiv c_{ji}$
from a site $S_i$ to its neighboring one $S_j$~\cite{Alavaetal}.  In
the case the external field is applied, only the local sum of fields,
$H+h_i$, is added to a spin towards the direction it is positive. The
graph-theoretical combinatorial optimization algorithms, namely
maximum-flow minimum-cut algorithms, enable us to find the bottleneck,
which restricts the amount of the flow which is possible to get from
the source to the sink through the capacities, of such a random graph.
This bottleneck, path $P$ which divides the system in two parts: sites
connected to the sink and sites connected to the source, is the global
minimum cut of the graph and the sum of the capacities belonging to
the cut $\sum_P c_{ij}$ equals the maximum flow, and is smaller than
of any other path cutting the system. The value of the maximum flow
gives the total minimum energy of the system.  The maximum flow
algorithms are proven to give the exact minimum cut of all the random
graphs, in which the capacities are positive and with a single source
and sink~\cite{network}. In physical situations this means the systems
are without local frustration. The algorithm was actually used for the
first time in this context by Ogielski~\cite{Ogielski}, who showed
that the 3D RFIM has a ferromagnetic phase. The best known maximum
flow method is by Ford and Fulkerson and called the augmenting path
method~\cite{FordF}. We have used a more sophisticated method called
push-and-relabel by Goldberg and Tarjan~\cite{Goltar88}, which we have
optimized for our purposes. It scales almost linearly, ${\mathcal
O}(n^{1.2})$, with the number of spins and gives the ground state in
about minute for a million of spins in a workstation.

When we have added an external field in the systems our system sizes
are restricted to $L^2 =175^2$, for $H$ small but nonzero due to the
range of integer variables (for numerical reasons we use a discrete
representation of real fields). When the high precision for field
values is not needed the computations extend up to system sizes $L^2
=1000^2$. We have used periodic boundary conditions in all of the
cases. Also the percolation is tested in the periodical or cylindrical
way, i.e., a cluster has to meet itself when crossing a boundary in
order to span a system.

When the red clusters are studied, Section~\ref{red_clust}, we have
applied a technique, which allows us to take advantage of the so
called {\em residual} graph~\cite{Ahuja}.  After the original ground
state is searched a perturbation is applied. This means that {\it
e.g.}, a spin is forced to be to the other direction with a large
opposite field value. Then the ground state is searched again. This
time all the flow need not to be constructed from scratch, but instead
one can utilize the final situation of the first ground state search
(the residual graph).  Only the extra amount of flow, needed since the
capacity of the large opposite field value is added, has to be forced
through the system to the sink. One has to subtract also the flow from
the original field value (retrace it back to the source). It is thus
convenient to reverse only the fields which originally were negative.
For the positive field values one would have to study a mirror copy of
the system ($h_i \rightarrow -h_i$). Thus we have analyzed the red
clusters only from the spanning clusters of down spins, which does not
disturb the statistics, since the spin directions are symmetrical. The
use of the residual graph reduces the time to calculate the next
ground state considerably, although approximately a half of the
spanning clusters have to be neglected.  Notice that since the
ground state energy is a linear function of the capacity of the
saturated bonds (or the field of the spins aligned along the local
field) one can compute the ``break-point'' field $h_b$, at which a
change takes place from the original ground state to the new one. We
have not paid attention to this, however, due to our main interest in
the geometry of the red clusters.  One interesting additional question
would be what is the smallest $h_b$ and its disorder-averaged
distribution.

\section{Destruction of ferromagnetic order}\label{breakup}

In this section we will derive the scaling for the break-up length
scale $L_b$, Eq.~(\ref{eq:L_b}), from extreme statistics (as done in
the paper by Emig and Nattermann~\cite{emig}). and confirm it with
exact ground state calculations. We also discuss the ensuing domain
structure qualitatively.

If one picks a (compact) subregion of area $a$ of a ferromagnetic 2D
RF system the energy is drawn from a Gaussian distribution
\begin{equation}
{\mathcal P}(E) = \frac{1}{\sqrt{2 \pi} \sigma} \exp\left\{- \frac{(E - 
\langle E \rangle)^2}{2 \sigma^2}\right\},
\label{Gauss}
\end{equation}
where the variance $\sigma^2 = a \Delta^2$ is due to the fluctuations
of random fields, and $\langle E \rangle \sim a$.  For a system of
size $L^2$ we have $N_a \sim L^2$ ways of making such a subregion.
The probability that a subregion has the lowest energy $E$ is given by
\begin{equation}
L_{N_a}(E) = N_a {\mathcal P}(E) \left \{1-C_1(E) \right \}^{N_a-1},
\label{global}
\end{equation}
where $C_1(E) = \int_{-\infty}^E P(\epsilon) \,
d\epsilon$~\cite{Galambos}.  The distribution $L_{N_a}(E)$ is in fact
a Gumbel distribution~\cite{Gumbel}.  The average value of the lowest
energies is given by
\begin{equation}
\langle E_0 \rangle = \int_{-\infty}^{\infty} E L_{N_a}(E) \,dE,
\end{equation}
which can not be solved analytically.  The typical value of the lowest
energy follows from an {\it extreme scaling} estimate.  The factor
inside the curly brackets in (\ref{global}) is close to unity if $C_1$
becomes small enough (for similar applications,
see~\cite{valley,periodic,CW_long,DBL}).  Thus \begin{equation}
\sigma N_a {\mathcal P}(\langle E_0 \rangle) \approx 1
\label{estimate}
\end{equation}
which yields,
\begin{equation}
\langle E_0 \rangle \approx \langle E \rangle - \sigma 
\{ \ln(N_a) \}^{1/2},
\label{typicalene}
\end{equation}
i.e. the energy gain from the fluctuations is
\begin{equation}
\langle E_g \rangle \approx \sigma  \{\ln(N_a)\}^{1/2}. 
\label{gain}
\end{equation}

\begin{figure}[f]
\centerline{\epsfig{file=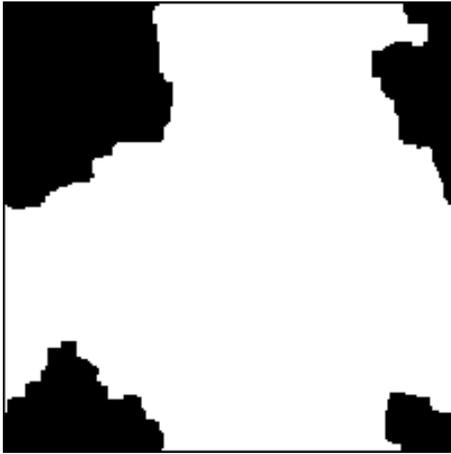,width=6cm}}
\caption{An example of the ground state after the first excitation,
$L^2 = 200^2$, Gaussian disorder, $\Delta = 0.76$. ``Up'' spins are
drawn in white, and ``down' spins in black. Note, that the system has
periodic boundaries.}
\label{fig1}
\end{figure}

A FM system would tend to take advantage of such large favorable
energy fluctuations by reversing a domain, which requires breaking
bonds. This is assumed to have a cost of
\begin{equation}
E_b \sim J a^{(d-1)/d}.
\label{cost}
\end{equation}
Equating Eqs. (\ref{gain}) and (\ref{cost}) yields the estimate of the
parameter values at which the first ``Imry-Ma'' domain occurs,
\begin{equation}
\sqrt{2 a}\Delta \{\ln(N_a)\}^{1/2} \sim Ja^{(d-1)/d}.
\label{Imryjump}
\end{equation}
It can be easily understood that the most preferable domain is the
one, which maximizes the area and minimizes the bonds to be broken,
which gives $a \simeq L^2/2$. Fig.~\ref{fig1} illustrates this, as we
increase (with a fixed random field configuration and system size) the
strength of the randomness or decrease the ferromagnetic couplings
until the first domain appears. It turns out to be so that the droplet
is of the order of the system size.  This kind of nucleation with a
critical size is reminiscent of a first order transition, and is
related to a level-crossing, when either the random field strength or
the system size is varied, similarly to random elastic manifolds, when
an extra periodic potential~\cite{periodic} or a constant external 
field~\cite{CW_long} is applied.  By substituting $a \sim L^2$ and
$N_a \sim L^2$ to (\ref{Imryjump}), we get for the length scale
\begin{equation}
L \sim \exp \left[A (J/\Delta)^2\right],
\label{eq:L_b_2}
\end{equation}
which is in fact $L_b$ as in (\ref{eq:L_b}). This result,
(\ref{eq:L_b_2}), is surprising in the sense, that the extreme
statistics calculation for the formation of a domain leads to the
exactly same scaling as the optimization of a domain wall energy on
successive scales in Binder's argument.

Due to the extensive size of the first domain-like excitation the
destruction of the ferromagnetism resembles a first order transition.
The magnetization for a certain disorder strength and system size
would be averaged over systems, in which the excitation has and has
not been formed yet, with $|m| \simeq 0$ and $|m| \simeq 1$,
respectively.  Hence we define a simpler measure for the break up of
FM order: the probability of finding a purely ferromagnetic system,
$P_{FM}(L,\Delta)$, i.e., for a fixed random field strength and system
size we calculate the probability over several realizations that
magnetization $|m| =1$~\cite{Oma98}. If the transition to the PM state
would be continuous, this would not make much sense, since already
small fluctuations would cause $P_{FM}(L,\Delta) \simeq 0$. However,
due to the first order behavior, and to the fact that the smallest
energy needed to flip a domain causes the excitation to be large,
$P_{FM}$ is a good measure and has a smooth behavior. We have checked
that $|m|$ vs. $P_{FM}$ does not depend on $L$.

\begin{figure}[f]
\vspace*{0.5cm} \centerline{\epsfig{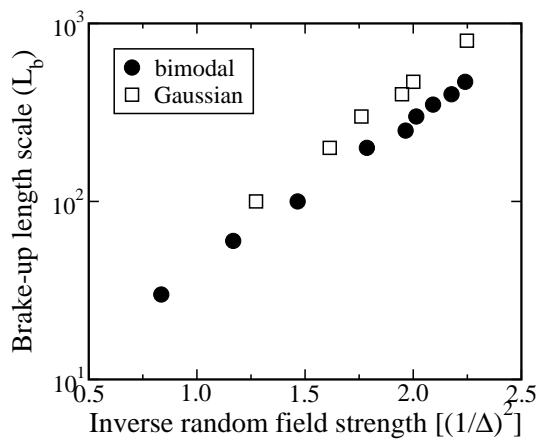}}\vspace*{0.5cm} 
\caption{The break-up length scale $L_b$ versus inverse random field
strength $(1/\Delta)^2$ for bimodal and Gaussian disorder (filled
circles and empty squares, respectively), calculated from $P_{FM}(L_b
) = 0.5$.}
\label{fig2}
\end{figure}

We have derived the break-up length scale $L_b$ by varying the random
field strength $\Delta$ from the probability of finding a pure
ferromagnetic system as $P_{FM}(L_b,\Delta) =0.5$. The data is shown
in Fig.~\ref{fig2} for Gaussian and bimodal disorder (in both cases
$J=1$), and the exponential scaling for $L_b$ vs. inverse random field
strength squared is clearly seen.  The prefactors are $A=2.1\pm 0.2$
and $1.9\pm 0.2$ for Gaussian and bimodal disorder, respectively.  To
check that the probability $P_{FM}(L,\Delta) < 1$ is not due to so
called stiff spins, i.e., single spins for which $h_i \geq 4J$, we
next derive an extreme statistics formula for their existence. The
probability of finding $h_i \geq 4J$ $P(h_i \geq 4J) = {\rm
erfc}(4/\Delta)$. The extreme statistics argument, $NP(h_i
\geq 4J) \approx 1$, with $N=L^d$, gives
\begin{equation}
L_1 \approx [{\rm erfc}(4/\Delta)]^{-1/2}.
\label{eq:L_1}
\end{equation}
For Gaussian $\Delta \simeq 0.8862$ for which $L_b =100$ in
Fig.~\ref{fig2} $L_1 \simeq 400$ from Eq.~(\ref{eq:L_1}). $L_1$ also
grows much faster than $L_b$ for decreasing $\Delta$ which is both
easy to see from Eq.~(\ref{eq:L_1}) and to check numerically. For
$\Delta \simeq 0.6670$ for which $L_b =800$ $L_1$ becomes as huge as
22300. To confirm further that the origin for the break-up is a large
domain, one can extend the argument to small domains. The length scale
$L_{2}$ at which one is able to find a cluster of two neighboring
spins flipped, i.e., $N/2 \, L_{2}^2 P(h_1 + h_2 \geq 6J) \approx 1$,
where $h_1 < 4J$, $6J-h_1 \leq h_2 \leq h_1$, becomes even greater
than $L_1$.  These small clusters are present in large system sizes,
but do not play a role in the probability of first excitation, since
the energy minimization prefers extensive domains. It is amusing to
note that the critical droplet size reminds of critical nucleation in
ordinary first order phase transitions. It is also worth pointing out
that the reasoning for stiff spins does not work for the bimodal
distribution (since it is bounded), and indeed we observe as expected
a similar $L_b$ scaling for both the Gaussian and bimodal disorders.

\begin{figure}[f]
\centerline{\epsfig{file=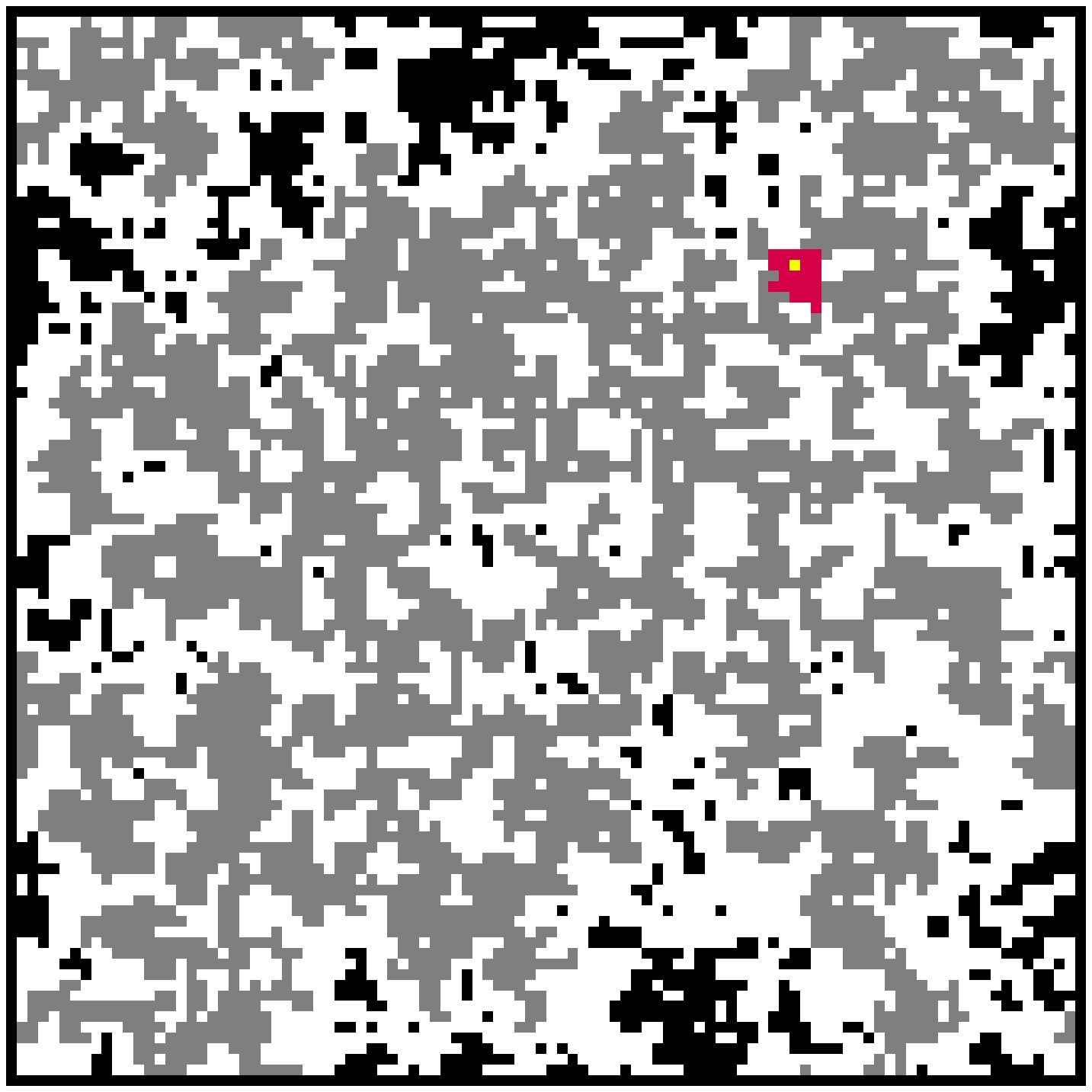,width=6cm}}
\centerline{\epsfig{file=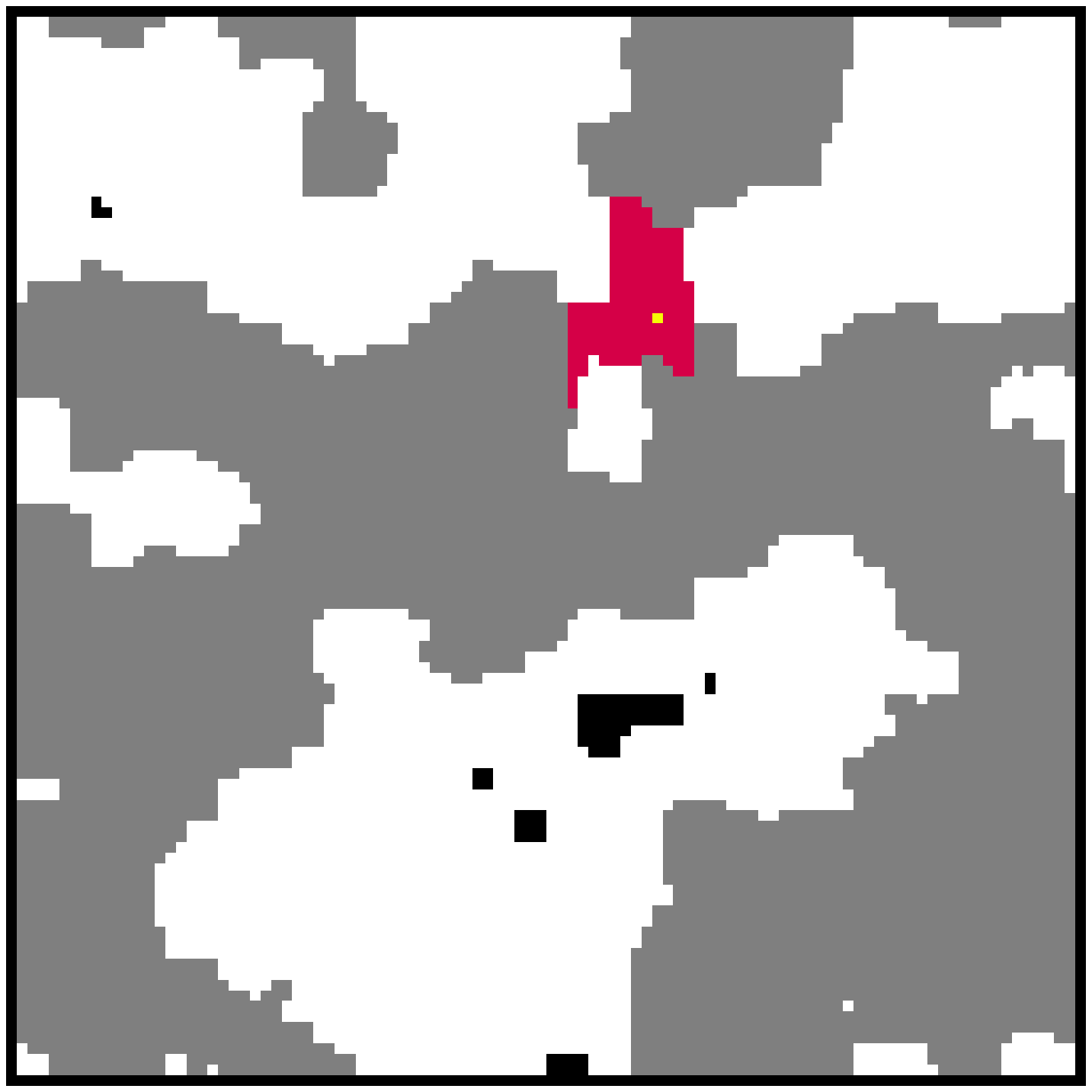,width=6cm}}
\caption{(color) Two examples of the ground state. ``Up'' spins 
in isolated domains are drawn in white, and ``down'' spins in black
unless they belong to the spanning cluster (grey). The yellow spin is
a seed of a red cluster, that breaks the spanning cluster.  Periodic
boundary conditions are used, and the spanning is checked in the
vertical direction. The system size $L^2=100^2$ and the random fields'
standard deviations $\Delta = 3.0$ and $\Delta = 1.2$.}
\label{fig3}
\end{figure}

When a system size is well above the break-up length scale the Imry-Ma
argument is no longer applicable to the structure.  In Fig.~\ref{fig3}
we depict two systems with a large Gaussian random field strength
value $\Delta =3.0$ and with a smaller one $\Delta =1.2$ for a system
size $L^2 = 100^2$. One can see that a system breaks to smaller and
smaller domains inside each other from the case seen in
Fig.~\ref{fig1}. The feature of having clusters in different scales is
familiar from the percolation problem~\cite{Amnon}. In fact in both of
the examples in Fig.~\ref{fig3} there is a domain, which spans the
system in vertical direction, drawn in gray. For the stronger random
field value one can see also smaller domains of different
sizes. However, the width of the spanning cluster is greater than in a
standard site or bond occupation percolation problem.  Later in
subsection \ref{perc-clust} we discuss the scaling properties of the
largest clusters in the ground state, above $L_b$.

\section{Magnetization and susceptibility with an external field}\label{suskis}

\begin{figure}[f]
\centerline{\epsfig{file=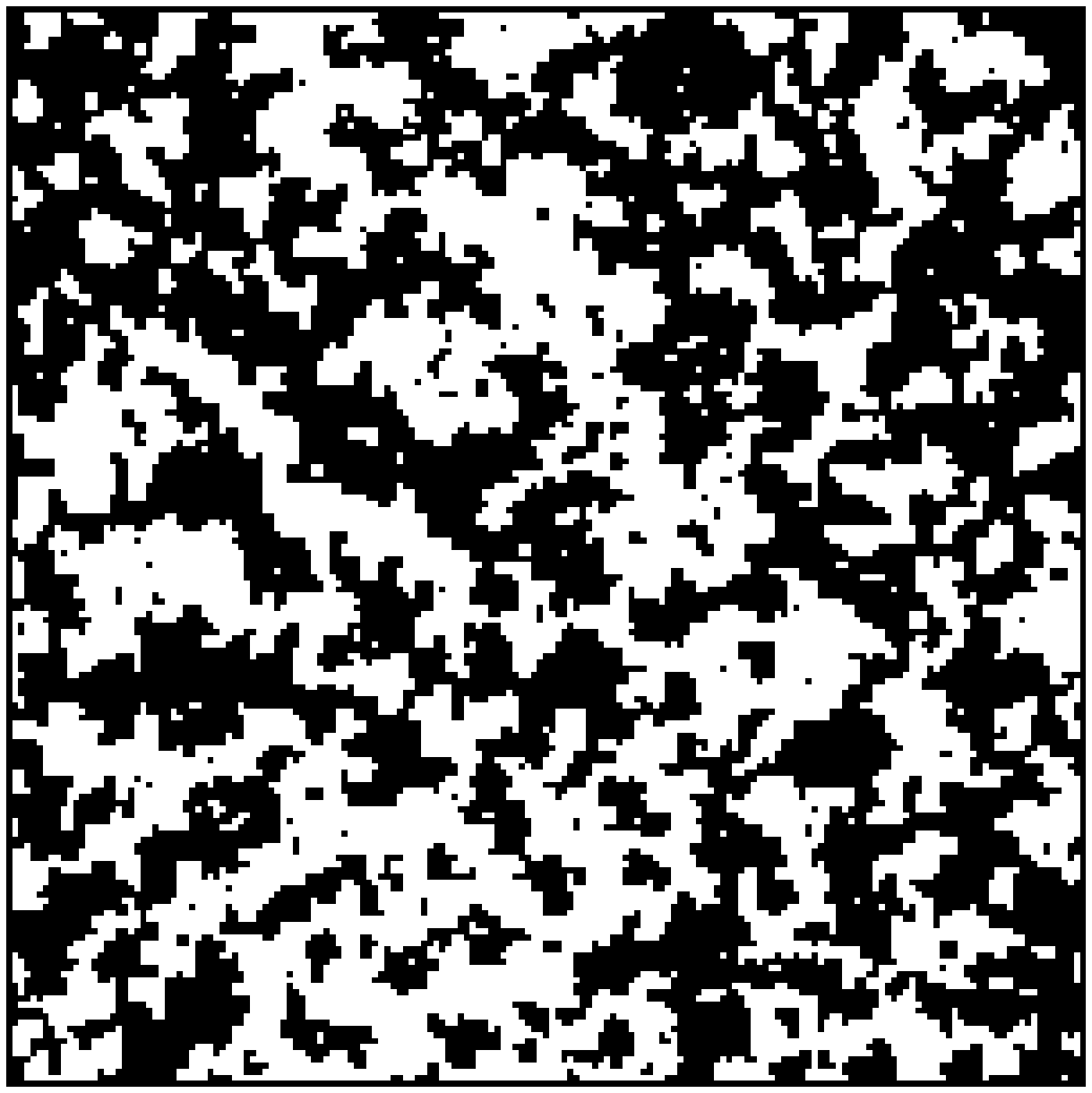,width=43mm}\epsfig{file=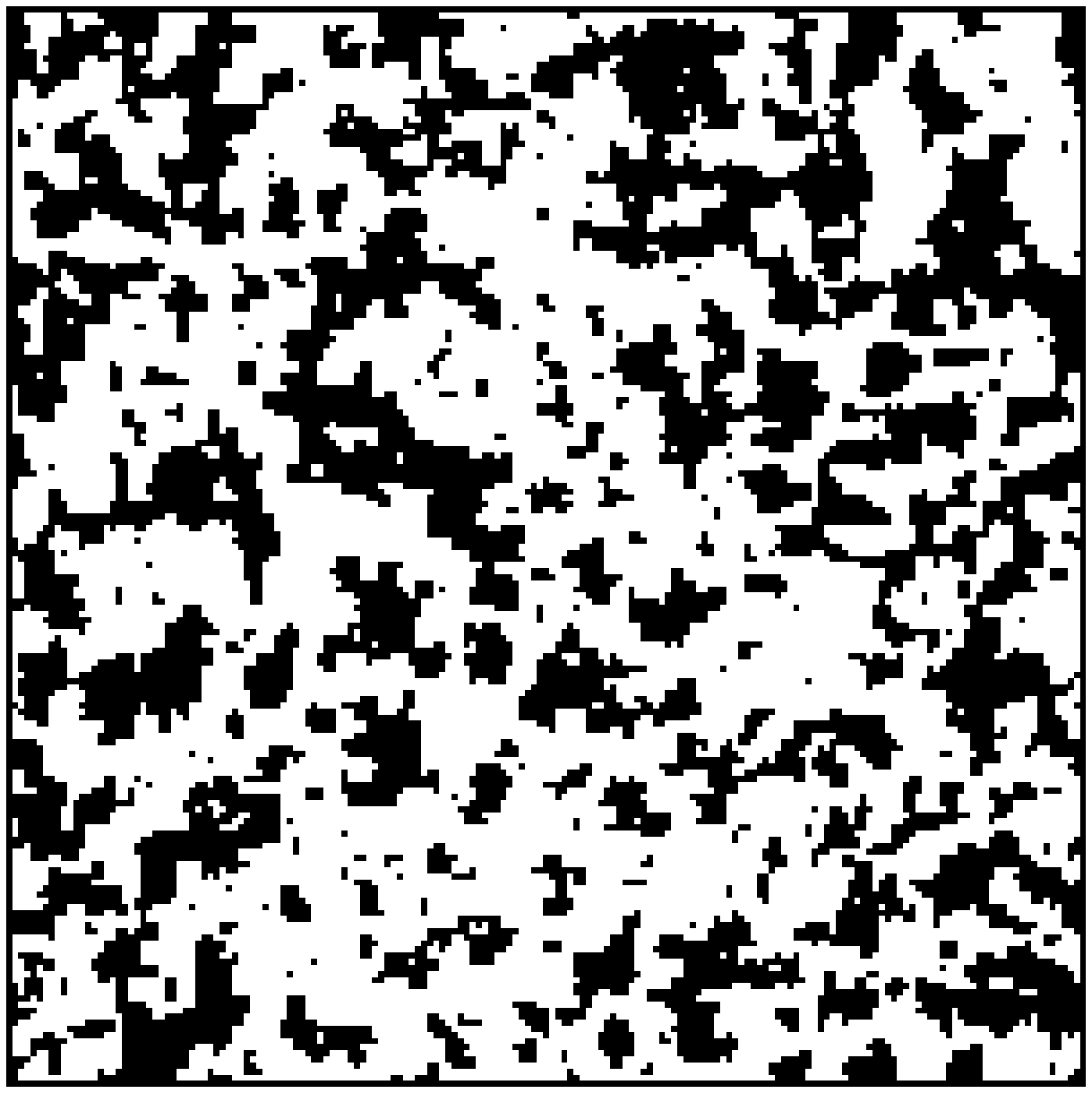,width=43mm}}
\centerline{\epsfig{file=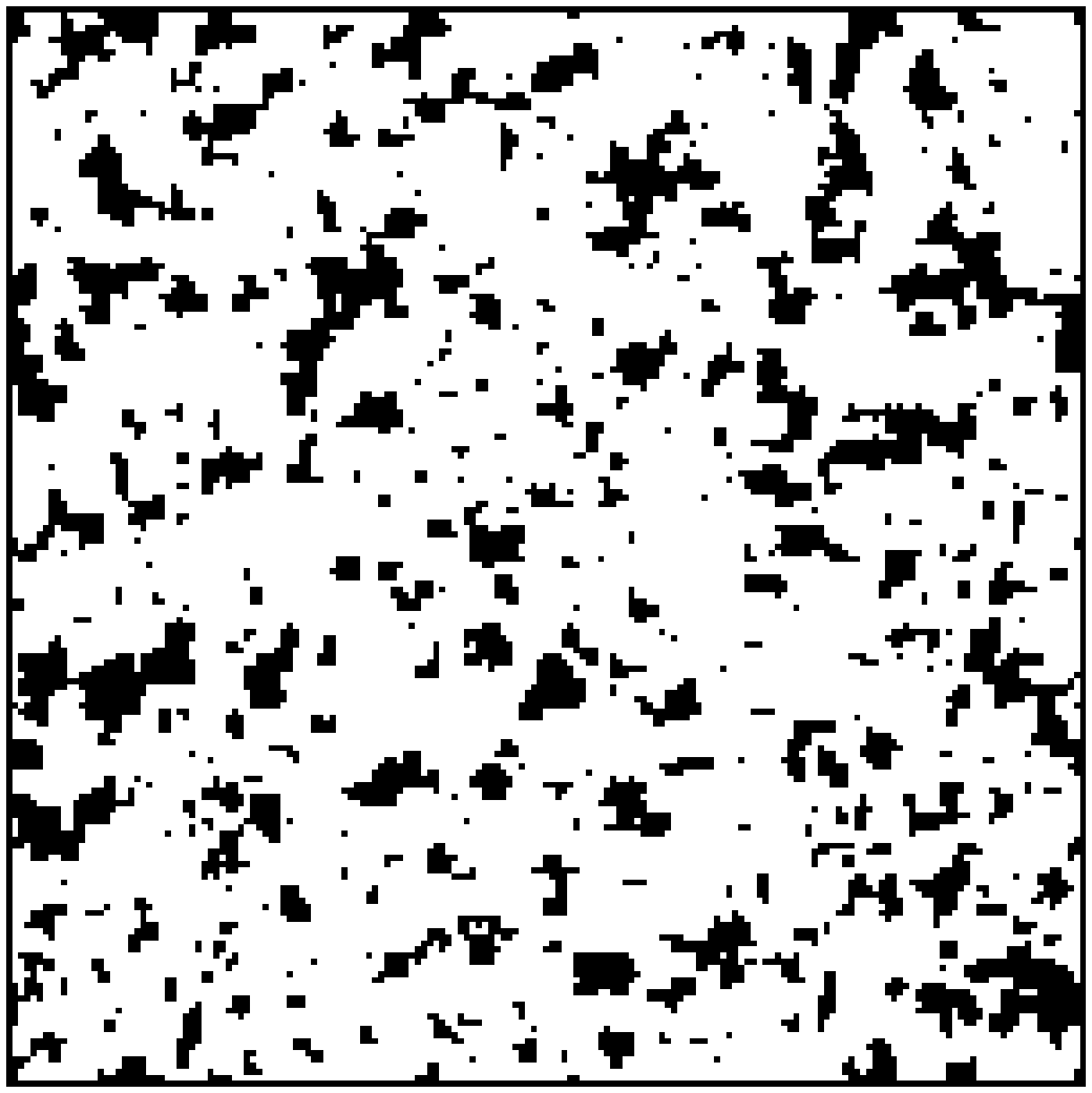,width=43mm}\epsfig{file=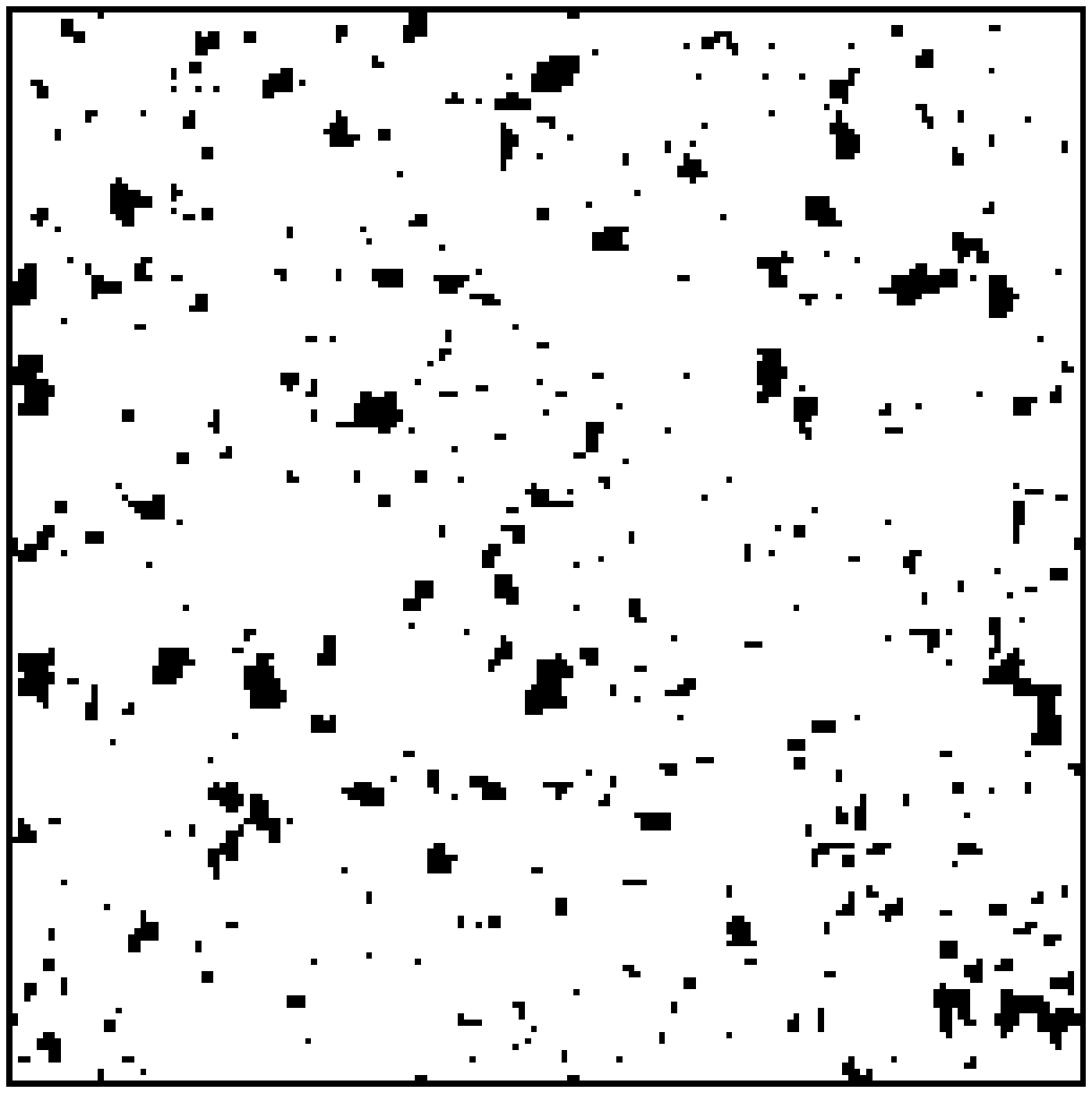,width=43mm}}
\caption{An example of the ground state, when an external field
is applied.  The system size $L =175 >L_b$. Gaussian disorder, $\Delta
= 1.9$. ``Up'' spins are drawn in white, and ``down'' spins in
black. The external field $H=$0.0 (l.h.s. up), 0.1 (r.h.s. up), 0.25
(l.h.s. down), and 0.5 (r.h.s. down).}
\label{fig4}
\end{figure}

\begin{figure}[f]
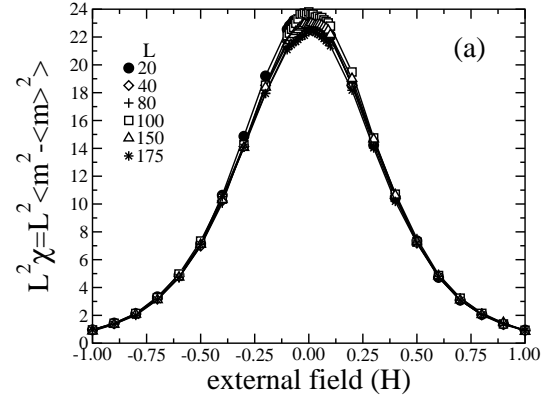

\vspace*{0.5cm} \centerline{\epsfig{file=fig5a.eps,width=7cm}}
\vspace*{1cm} \centerline{\epsfig{file=fig5b.eps,width=7cm}}
\vspace*{1cm} \centerline{\epsfig{file=fig5c.eps,width=7cm}} \vspace*{0.5cm} 
\caption{(a) The magnetization $m$ versus the external field  $H$
for the system size $L^2=175^2$ (all the tested system sizes $L>L_b$
lie exactly on top of each other), and random field strength values
$\Delta =$ 1.9, 2.0, 2.2, 2.6, 3.0, 3.5 and 4.5.  Each point is a
disorder-average over 5000 realizations and the error bars are smaller
than the symbols.  (b) The external field values $H_m$, when the
magnetization curves in (a) crosses the fixed magnetization values
$m=$ 0.25, 0.2, 0.1, -0.1, -0.2, -0.25, versus random field
strength. The $\exp$-lines are guides to the eye, and their prefactors
are estimated using least-squares fit.  (c) The data-collapse of (a)
by scaling $H$ with $\exp(-6.5/\Delta)$ estimated in (b).}
\label{fig5}
\end{figure}

In Fig.~\ref{fig4} we show what happens in system, with a system size
well above $L_b$ when an external field $H$ is applied.  Now the
clusters ``melt'' smoothly when the external field strength is
increased and such first order type of phenomenon is not seen as when
a first Imry-Ma droplet appears in the zero-field case.  The
magnetization has a continuous behavior, see Fig.~\ref{fig5}(a), where
we have the magnetization with respect to the external field for
several Gaussian disorder strength values. All the magnetization
values for different system sizes lie exactly on top of each other,
when $L > L_b$, and as long as the statistics is good.

For smaller system sizes $L<L_b$ one could study ``avalanche''-like
behavior, i.e. the sizes of the areas that get turned with
the magnetic field (see \cite{Frontera}). However, these are due to the first
order break-up, defined by (\ref{eq:Hc_binder}) and one should bear in
mind that such behavior does not exist in the thermodynamic limit, $L
\to \infty$, when the system sizes are above the break-up length
scale. For $L>L_b$ our results indicate that the size distribution
of the flipped regions as $H$ is swept is not that interesting.

In order to find the scaling between the external field and the random
field strength we have taken from Fig.~\ref{fig5}(a) the crossing
points of magnetization curves with a fixed magnetization values at
external fields $H_m$ for different random field strength values
$\Delta$. The external field $H_m$ scales exponentially with respect
to the random field strength,
\begin{equation}
H_m \sim \exp(-6.5/\Delta), 
\label{Eq:H_D_magn}
\end{equation}
see Fig.~\ref{fig5}(b).  This is also an evidence of {\em
non-existence} of a critical point in $\Delta$, in which case there
should be a power-law behavior if the transition was continuous, thus
no PM to FM transition is seen. The data-collapse using the scaling
(\ref{Eq:H_D_magn}) is shown in Fig.~\ref{fig5}(c) confirming the
prediction of the scaling. The magnetization has a linear behavior
with respect to the external field for small field values $H$ and
exponential tails. The exponential behavior of Eq.~(\ref{Eq:H_D_magn})
implies that there is a unique ``melting rate'' at which the cluster
boundaries get eroded as $H$ increases and that the process is
otherwise similar for all $\Delta$. We have no analytical argument for
the melting rate [or the slope of the $m(H)$-curve], and note that it
is not seemingly, at least, related to $L_b$.

We have also studied the susceptibility, $\chi \sim \langle m^2 -
\langle m \rangle^2 \rangle$, with respect to the external field.
In Fig.~\ref{fig6}(a) the susceptibility is shown for a fixed random
field strength $\Delta = 2.2$ and varying system size.  The data has
been collapsed with the area of the systems, $\chi/L^{-2}$.  In
Fig.~\ref{fig6}(b) we have data-collapsed the susceptibility versus
random field strength by scaling the external field with
(\ref{Eq:H_D_magn}) as for magnetization and the susceptibility with
$\chi \sim \exp(7.3/\Delta)$.  Again the exponential behavior is a
sign of non-existence of any critical point, due to the lack of
power-law divergence at any $\Delta_c$. Although the shape of the
data-collapse of the susceptibility looks almost Gaussian, it is
actually not. It has a constant value for small external field values
$H$ and exponential tails for large values, as seen in
Fig.~\ref{fig6}(c). This results straightforwardly from the
magnetization, since $\chi = \partial m/\partial H$. To summarize the
behavior of the susceptibility, it is
\begin{equation}
\chi \sim L^{-2} \exp(7.3/\Delta) g(H/H_m),
\label{Eq:chi}
\end{equation}
where $H_m$ is from Eq.~(\ref{Eq:H_D_magn}) and 
\begin{equation}
g(y) \sim \left\{ \begin{array}{lll}
{\rm const},&\mbox{\hspace{5mm}}&y \simeq 0, \\
\exp(-0.2|y|),& &y \to \pm \infty. 
\end{array} \right.
\label{Eq:chi_g}
\end{equation}
Therefore the fluctuations of the magnetization are associated with
yet another scale, which is almost but not quite an inverse of that
related to the magnetization.  From the suscpetibility one gets the
magnetization correlation length $\xi_{m}$, which has an
exponential dependence on the random field strength. It should be
noted, finally, that we have here studied only the case with Gaussian
disorder. With any other distribution we would expect that the
prefactors in Eqs.~(\ref{Eq:H_D_magn}), (\ref{Eq:chi}), and
(\ref{Eq:chi_g}) would change.

\section{Percolation with an external field}\label{perkolaatio_H}

Motivated by Fig.~\ref{fig3}, where the domains resemble the
percolation problem we next study the percolation behavior in the 2D
random field Ising magnets with Gaussian disorder. The usual bimodal
distribution could be studied as well, but since it is susceptible to
some anomalous features we concentrate on the Gaussian case which does
not have these problems. The bimodal case suffers from the fact
that the ground states are highly degenerate at fractional field
strength values. Thus there are some ambiguities in how define
percolation clusters~\cite{Alexander,Sorin}.

When the random field strength is well above the coupling constant
value, $\Delta \gg J$, the percolation can be easily understood by
considering it as an ordinary site-occupation problem. This means
that, only the random field directions are important and the coupling
constants may be neglected.  The site-percolation occupation threshold
probability for square-lattices is $p_c \simeq 0.593$~\cite{Amnon},
i.e., well above one half. Applied to the strong random field strength
case it means that there must be a finite external field in order to
get a domain spanning the system. However, when the random field
strength is decreased, the coupling constants start to contribute and
in some cases systems span even without an external field, as in
Fig.~\ref{fig3}.  Hence, we propose a phase-diagram Fig.~\ref{fig7}.
There we can take the limit $1/\Delta =0$ so
that the ordinary site-percolation problem is encountered. This is
true for distributions for which one can control the fraction of
``stiff'' spins (i.e. $h_i > 4J$) systematically. In the case of
Gaussian disorder there will be of course, even for $\Delta$ very
large a small fraction of ``soft'' spins where this criterion is not
fulfilled. Thus the exact point that the percolation line approaches
in the $1/\Delta\rightarrow 0$-limit will depend on the distribution,
but we expect that the ``$p_c$'' is different from one-half, that is
$H_c \not= 0$. Notice that again the binary distribution presents a
problem.

When $1/\Delta \to \infty$ the percolation threshold lines start to
approach each other and the $H=0$ line. Now there arises two
questions. The first one is, what kind of a transition is the
percolation here?  Is it like the ordinary short-range correlated
percolation, suggested by the site-percolation analogy for the strong
random field strength case, or are there extra correlations due to the
global optimization relevant here?  Examples about similar cases can
be found from Ref.~\cite{corrperc}.  The second question is, do the
lines meet each other at finite $\Delta_c$, i.e. does there exist a
spanning cluster also when $H=0$ and $\Delta >0$? Our aim is to answer
these two questions in this section, where we study the percolation
problem in the vertical direction in the phase-diagram
Fig.~\ref{fig7}, and in the next section, where the horizontal
direction, $H=0$ line, is considered.

In Fig.~\ref{fig8}(a) we have drawn the spanning probabilities of up
spins $\Pi_{up}$ with respect to the external field $H$ for several
system sizes $L$, which are greater than $L_b$, for a fixed random
field strength $\Delta = 2.6$. The curves look rather similar to the
standard percolation. When we take the crossing points $H_c(L)$ of the
spanning probability curves with fixed spanning probability values for
each systems size $L$, we get an estimate for the critical external
field $H_c$ using finite size scaling, see Fig.~\ref{fig8}(b). There
we have attempted successfully to find the value for $H_c$ using the
standard short-range correlated 2D percolation correlation length
exponent $\nu=4/3$. Using the estimated $H_c = 0.00186$ for
$\Delta=2.6$ we show a data-collapse of $\Pi_{up}$ versus
$(H-H_c)/L^{-1/\nu}$ in Fig.~\ref{fig8}(c), which confirms the
estimates of $H_c$ and $\nu=4/3$~\cite{Amnon}. We get similar
data-collapses for various other random field strength values $\Delta$
as well.  In order to test further the universality class of the
percolation transition studied here, we have also calculated the order
parameter of the percolation, the probability of belonging to the
up-spin spanning cluster $P_\infty$. Using the scaling analysis for
the correlation length
\begin{equation}
\xi_{perc} \sim |H-H_c|^{-\nu},
\label{xi_perc}
\end{equation}
and for the order parameter, when $L < \xi_{perc}$,
\begin{equation}
P_\infty(H) \sim (H-H_c)^{\beta},
\label{beta}
\end{equation}
we get the limiting behaviors,
\begin{equation}
P_\infty(H,L) \sim \left\{ \begin{array}{lll}
(H-H_c)^{\beta} &\mbox{\hspace{5mm}}& L < \xi_{perc},\\
L^{-\beta/\nu}& &L > \xi_{perc}, 
\end{array} \right.
\label{P_infty_limit}
\end{equation}
and thus the scaling behavior for the order parameter becomes
\begin{eqnarray}
P_\infty(H,L) \sim L^{-\beta/\nu} F\left[\frac{(H-H_c)^{-\nu}}{L}\right] 
\nonumber \\
\sim L^{-\beta/\nu} f\left(\frac{H-H_c}{L^{-1/\nu}}\right).
\label{P_infty_scale}
\end{eqnarray}
We have done successful data-collapses, i.e. plotted the scaling
function $f$, for various $\Delta$ using the standard 2D short-range
correlated percolation exponents $\beta=5/36$ and $\nu =4/3$, of which
the case $\Delta =3.0$ with $H_c=0.040$ is shown in
Fig.~\ref{fig9}. Note, that the values $(H-H_c)/L^{-1/\nu} > 0$ are
not shown, since cut-offs appear, due to the fact that $P_\infty$ is
bounded in between $[0,1]$ and the non-scaled $P_\infty =1$ values
after scaling saturate at different levels depending on the system
size.

Hence, we conclude that the percolation transition for a fixed
$\Delta$ versus the external field $H$ is in the standard 2D
short-range correlated percolation universality
class~\cite{Amnon}. This is confirmed by the fractal dimension of the
spanning cluster, too, as discussed below. Also other exponents could
be measured, as $\gamma$ for the average size $\langle s \rangle $ of
the clusters, and $\sigma$ and $\tau$ for the cluster size
distribution.  Note, however, that then the control parameter should
be the external field $H$ instead of the disorder strength
$\Delta$. See Ref.~\cite{Ess97} for an example of the cluster size
distribution for a non-critical case ($H=0$, but $|H_c| \gg 0$). The
other exponents should be measurable, too, like the fractal dimension
of the backbone of the spanning cluster, the fractal dimension of the
chemical distance, the hull exponent etc.  In addition to the correct
control parameter also the break-up length scale, has to be
considered, too.  Notice that there is an slight contradiction hidden
in the notion that the hull exponent could be measured. Namely, both
the work of Ref.~\cite{Ess97} and studies of domain walls enforced
with appropriate boundary conditions give no evidence thereof.  It
seems likely that to recover the right exponent (4/3) one has to
resort to studying the spanning cluster geometry itself at the
critical point with $L > L_b$. It is amusing to note that the standard
2D percolation hull exponent can be recovered in non-equilibrium
simulations of 2D RFIM domain walls~\cite{domwalls}.

We have now shown, that there exists a line of critical external field
values $H_c(\Delta)$ for percolation.  The corresponding correlation
length $\xi_{perc}$ diverges as (\ref{xi_perc}) with a correlation
length exponent $\nu=4/3$. On the other hand it was shown in the
previous section, that there is no critical external field value for
the magnetic behavior, i.e., no PM to FM transition, and the magnetic
correlation length $\xi_m$ has an exponential dependence on
$\Delta$. The percolation correlation length $\xi_{perc}$ may cause
some confusion, when studying the PM structure of the GS, since it
brings in another length scale.

To answer the question how the percolation critical external field
$H_c(\Delta)$ behaves with respect to the random field, we have
attempted a critical type of scaling using the calculated $H_c$ for
various $\Delta =$2.0, 2.2, 2.6, 3.0, and 4.5. For smaller $\Delta$
$L_b$ becomes large and $H_c$ approaches the vicinity of zero, being
thus numerically difficult to define. We have been able to use the
Ansatz behavior of
\begin{equation}
H_c \sim (\Delta -\Delta_c)^\delta,
\label{Eq:H_c_D_c}
\end{equation} 
where $\delta=2.05\pm 0.10$.  In Fig.~\ref{fig10}(a) we have plotted
the calculated $\Delta$ values versus the scaled critical external
field $[H_c(\Delta)]^{1/2.05}$ and it gives the estimate for $\Delta_c
= 1.65 \pm 0.05$. This indicates that the percolation probability
lines for up and down spins meet at $\Delta_c = 1.65$ and for $\Delta$
below the critical $\Delta_c$ there is always in the systems spanning
of either of the spin directions, even for $H=0$. Actually one should
note, that the only way that neither of the spin directions span is to
have a so called checker-board situation, which prevent both of the
spin directions to have neighbors with the same spin
orientation. However, the another scenario with an exponential
behavior for $H_c(\Delta)$ fits also reasonably well. This would
suggest, that there is no finite $\Delta_c$. Fig.~\ref{fig10}(b) shows
a behavior of $H_c \sim \Delta^2\exp (-13/\Delta^2 -4)$.  This can be
compared with Eq.~(\ref{eq:Hc_binder}), where the break-up external
field was derived. Notice that the derivation was for compact domains
and the spanning clusters here are by default fractals. Besides that,
the factor 13 in front of $1/\Delta^2$ is much larger than $A =
2.1$ in the scaling form for $L_b$.  The difference implies that the
$L_c$, at which length scale the spanning probability vanishes, scales
as $L_c \sim L_b^6$. The $L_b$ is already exponentially large length
scale for small $\Delta$, so $L_c$ should be large enough that one can
be below it in experiments, and thus a system can
``apparently percolate''~\cite{exptl,Belanger97}.

\section{Percolation at $H=0$}\label{perkolaatio_0}

To understand how the percolation transition is seen when there is no
external field and the random field strength is changed we study the
phase diagram in Section~\ref{span_perk_0} in the direction of the
horizontal arrow in Fig.~\ref{fig7}. The structure of the spanning
clusters are studied in Sections \ref{perc-clust} and in
\ref{red_clust} with the help of the so called {\em red clusters}.

\subsection{Spanning probability}\label{span_perk_0}

In Fig.~\ref{fig11}(a) we have plotted the probability for spanning of
either up or down spins $\Pi$ as a function of the Gaussian random
field strength $\Delta$. The probabilities are calculated up to
$\Delta =30$, but only the interesting part of the plot is shown.
There is a drop from $\Pi = 1$ at $\Delta$, which corresponds $L =
L_b$ for each system size, to a value about $\Pi \simeq 0.85$. We have
also calculated the $\Pi_{up}$ in this case and it is approximately
one half of $\Pi$. For the larger random field strength values the
probabilities $\Pi$ decrease and the lines get steeper, when the
system size increases. In order to see, if the spanning probabilities
are converging towards a step function at some threshold value, we
have calculated the probabilities up to the system size $L^2 =1000^2$
and each point with 5000 realizations.

For each system size $L$ we have searched the crossing points
$\Delta_c(L)$ of the spanning probability curves in
Fig.~\ref{fig11}(a) with fixed probability values $\Pi=$0.1, 0.15,
0.2, 0.25, 0.3, 0.4, and 0.5. Using finite size scaling for
$\Delta_c(L)$ of the form $\Delta_c(L) = \Delta_c (1+ C_1
L^{-1/\nu})(1+C_2 L^{-1/\nu_2})$ we have estimated $\Delta_c$ for each
$\Pi$ value, see Fig.~\ref{fig11}(b). There we have plotted the
$\Delta_c(L)$ versus the scaled system size $C_1 L^{-1/\nu}$.  One
sees that the different threshold values for spanning probabilities
$\Pi$ approach different critical random field strength values
$\Delta_c$. The threshold $\Pi$'s have been plotted with respect to
$\Delta_c$ in Fig.~\ref{fig11}(c).  In the ordinary percolation, this
should be a step function, and the correlation length exponent $\nu$
independent on the criterion $\Pi$. However, here also $1/\nu$ is
dependent on the criterion $\Pi$ and varies with respect to
$\Delta_c(\Pi)$. We believe, that this surprising phenomenon is due to
that we are approaching the part in the phase diagram Fig.~\ref{fig7},
where the percolation lines of up and down spins are getting close to
each other. In terms of the two control parameters $\Delta$ and $H$
one can think about the ``percolation manifold'': it has a line of
unstable fixed points $H_c(\Delta)$. Usually $H$ is a good control
parameter close to $H_c$. Having $\Delta$ as a control parameter seems
to have the problem that one moves almost parallel to the actual line
$H_c(\Delta)$.

When considering the percolation probability of up or down spins, it
actually consists of probabilities of up-spin spanning $\Pi_{up}$
and down-spin spanning $\Pi_{down}$ as $\Pi = \Pi_{up} + \Pi_{down} -
\Pi_{up}\Pi_{down}$, since they are correlated with each other.  
Assuming that $\Pi_{up}$ (and $\Pi_{down}$ respectively) has a value
about one-half at the critical line of percolation at the
thermodynamic limit, we get $\Pi = 0.75$. In standard percolation such
a value is not actually universal (and we have not confirmed it) but
depends on the boundary conditions, etc.~\cite{J-P}. However, whatever
the values for $\Pi_{up}$ and $\Pi_{down}$ are at the thermodynamic
limit, as long as they are below unity, $\Pi$ is below unity,
too. This may be the reason, why there is an immediate drop in
Fig.~\ref{fig11}(a) from $\Pi = 1$ at $\Delta(L_b)$ for each system
size, to a value about $\Pi \simeq 0.85$. If we approximate with a
linear behavior the $\Pi$ versus $\Delta_c(\Pi)$ in
Fig.~\ref{fig11}(c), the critical value estimated in the previous
section $\Delta_c = 1.65 \pm 0.05$, when the percolation threshold
$H_c =0$ for up-spin spanning, has a value about $\Pi \simeq
0.7$. Another interesting point in Fig.~\ref{fig11}(c) is that the
$1/\nu$ is about 3/4 , when $\Pi$ versus $\Delta_c(\Pi)$ approaches
zero. Thus the standard correlation length exponent would be reached
far enough away from the area, where the percolation threshold lines
for up and down spins touch each other.

In order to test the break-up length scale type of scaling for the
percolation behavior (\ref{eq:L_b}), we have taken from
Fig.~\ref{fig11}(a) the estimated $\Delta_c(L)$ for various $\Pi$
values and plotted the system sizes in double-logarithm scale versus
the logarithm of the inverse of the critical $\Delta_c(L)$, see
Fig.~\ref{fig11}(d). The exponent, which is $\alpha =2$ in $L_b$
scaling, $L_b \sim \exp(A/\Delta^\alpha)$, is now dependent on $\Pi$
again. At least this does not solve the problem here, and the break-up
length scale type of scaling can be ruled out.

\subsection{The percolation cluster}\label{perc-clust}

In order to see if the thickness of the spanning cluster affects the
scaling of the standard percolation we have measured the fractal
dimension of the spanning cluster when $H=0$. By now unsurprisingly,
the standard two-dimensional short-range correlated percolation
fractal dimension $D_f =91/48$ fits very well in the data, as can be
seen in Fig.~\ref{fig12}. The least-squares fit gives a value of $D_f
= 1.90 \pm 0.01$. We have measured also the sum of the random fields
in the spanning cluster and found that the sum scales with the
exponent $D_f =91/48$, too. This is in contrast to the Imry-Ma domain
argument, where the sum is taken scale as $L^{d/2}$.  The prefactor
for the scaling of the sum of the random fields approaches slowly zero
with decreasing random field strength, opposite to the mass of the
spanning cluster, which increases with decreasing $\Delta$.

Hence, the Imry-Ma argument defines only the {\em first} excitation,
and is irrelevant when it comes to domains when the system has broken
up to many clusters on different length scales.  Then the structure is
due to a more complicated optimization. The domains are no longer
compact and as noted above for large enough domains the domain-wall
length should be characterized by the percolation hull exponent.

\subsection{Red clusters}\label{red_clust}

So far all the evidence points out to the direction that the
percolation transition is exactly of the normal universality class. To
further investigate the nature of the clusters in the presence of the
correlations from the GS optimization, we next look at the so-called
{\em red clusters}.  The structure of a standard percolation cluster
can be characterized with the help of the ``colored sites'' picture in
which one assesses the role of an element to the connectivity of the
spanning cluster.  This picture has been also called the {\it
links-nodes-blobs} model with dead-ends~\cite{Amnon}. The red sites,
or links and nodes, are such that removing any single one breaks up
the spanning cluster.

To compare with the original ground state we investigate what happens
if one inverts, by fixing the local field $h_i$ to a large value
opposite to the spin orientation, any spin belonging to the spanning
cluster. Then the new GS is found with this change to the original
problem. The effect is illustrated in Fig.~\ref{fig3}, the crucial
difference to site-percolation is that now a whole sub-cluster can be
reversed.  The spin drawn in yellow is the inverted, ``seed'' spin in
the spanning cluster, and the spins painted red form the rest of the
red cluster, which are flipped from the original ground state when the
energy is minimized the second time.  We do the investigation whether
the original cluster retains its spanning property for each spin or
trial cluster in analogy with ordinary percolation. Those spins that
lead to a destructive (cluster) flip, define then {\em red clusters}
(RC) as all the spins that reversed simultaneously.

The finite size scaling of the number of the red clusters, $\langle
N_{RC} \rangle$, is shown in the Fig.~\ref{fig13}(a) for different
field values.  $\langle N_{RC}\rangle$ is in practice calculated as
the number of the seed spins, which cause the breaking of the spanning
cluster, since two different seed spins may both belong to the same
red clusters without the red clusters being identical. The technique
to find the red clusters was introduced in Section~\ref{method} and
although it is efficient, only up to the system size ${\mathcal
O}(200^2)$ can be studied, since each of the spins in spanning
clusters has to be checked separately, because one cannot know
beforehand, whether it is critical or not - this is what we want to
find out.  For smaller field values the spanning cluster is
``thicker'' and the red clusters get larger.  One can see from the
Fig.~\ref{fig13}(a) that $\langle N_{RC} \rangle $ scales with
$L^{1/\nu}$, where $\nu \simeq 4/3$ as in ordinary percolation, for
field values $\Delta \leq \Delta_c$, when $L > L_b$. The amplitude is
larger the smaller the field, as is the average mass of red clusters
$\langle M_{RC} \rangle$.  $\langle M_{RC} \rangle$ is independent of
the system size $L$ and depends only on the field $\Delta$, see
Fig.~\ref{fig13}(b).

The other elements of the spanning cluster, dead-ends and blobs, could
be generalized, too. Here blobs, which are multiple connected to the
rest of the spanning cluster, are such that in order to break a
spanning cluster, several seed spins are needed to flip simultaneously
instead of a single one. Links, nodes, and blobs form together the
backbone of the spanning cluster, and the rest of the mass of the
cluster is in the dead-ends. The red cluster size scale defines the
average smallest size of any element of the spanning cluster.

\section{Conclusions}\label{concl}

In this paper we have studied the character of the ground state of the
two-dimensional random field Ising magnet. We have shown that the
break-up of the ferromagnetism, when the system size increases, can be
understood with extreme statistics. This length scale has been
confirmed with exact ground state calculations.  The change of
magnetization at the droplet excitation is naturally of ``first-order''
-kind.

Above the break-up length scale we have studied the magnetization and
susceptibility with respect to a constant external field. The behavior
of the magnetization and the susceptibility is continuous and smooth
and does not have any indications of a transition or a critical point,
in agreement with the expectations of a continuously varying
magnetization around $H=0$, and a paramagnetic ground state.  We thus
conclude that the correct way of looking at the susceptibility is to
study it with respect to the external field and above the break-up
length scale instead of as a function of the random field strength,
when the first order character of the break-up length scale may among
others cause problems.

However, we are able to find another critical phenomenon in the
systems, in their geometry. For square lattices sites do not have a
spanning property in ordinary percolation, when the occupation
probability is one half.  This corresponds to the random field case
with high random field strength value without an external field. When
an external field is applied and the random field strength decreased,
a percolation transition can be seen. The transition is shown to be in
the standard 2D short-range correlated percolation universality class,
when studied as a function of the external field. Hence, the
correlations in the two-dimensional random field Ising magnets are
only of finite size.  We also want to point out that in these kind of
systems, the random field strength is a poor control parameter and the
systems should be studied with respect to the external field, and
after that map to the random field strength. By doing so we have been
able to find a critical random field strength value, below which the
systems are always spanning even without an external field. When the
percolation transition is studied without an external field and tuning
the random field strength lots difficulties are encountered. This
might cause puzzling consequences when studying the character of the
ground states, not only because of the bad control parameter, but also
because the percolation correlation length may be mistaken for
something as the magnetization correlation length. Also note that the
``true behavior'' is seen only for system sizes large enough ($L >
L_b$).

The percolation character of the ground state structure can be
measured by the standard percolation fractal dimensional scaling for
the mass of the spanning domain. The existence of such a large cluster
is not against the paramagnetic structure of the ground state, since
the fractal dimension is below the Euclidean dimension. In order to be
consistent with the Aizenman-Wehr argument in the zero-external field 
limit the spins in the opposite direction from the external field may
form the spanning cluster at low random field strength values. 
In fact we have found cases of finite systems
for $H=0$, where the magnetization of the system is opposite to the
orientation of spins in the spanning cluster. Notice that this
does not imply that the critical lines $H_c (\Delta)$ actually
cross each other at $\Delta_c$ continuing on the opposite 
side of the $H=0$-axis (see Fig.~\ref{fig7}). By considering
the red clusters it seems that in the TD-limit the spanning cluster 
should be broken up at $H = \epsilon$, $\epsilon \rightarrow 0$
since the field needed
to flip such a critical droplet should go to zero with $L$.
Also, since the sum of
the fields in the spanning cluster is shown to scale with the same
fractal dimension as the mass, we conclude that the Imry-Ma argument
does not work any more after the system has broken up on several
domains. It works only for the first domain to appear.

We have also generalized the red sites of the standard percolation to
red clusters in the percolation studied here. A red cluster results
from the energy minimization by flipping a whole cluster although only
a single spin has been forced to be flipped, and breaks up the
spanning character of a percolating cluster. Actually the finite size
of the red clusters indicate also the presence of only short-range
correlations in the systems. Such a lack of long-range correlations
maybe explains, why we can see an ``accidental'' percolation
phenomenon in a zero-temperature magnet whose physics is governed by
the disorder configuration.  The normal percolation universality class
is tightly connected to conformal invariance, which is most often
destroyed by long-range correlations or randomness \cite{CardyII}.

To finish the paper, we would like to raise some open questions
related to the percolation behavior of the ground states of the
two-dimensional random field Ising magnets. As noted, an interesting
problem is the exact relation of the RFIM percolation to conformal
invariance. The percolation characteristics of the ground state might
be experimentally measurable since the overlap of the ground state and
finite temperature magnetization should be close to unity for small
enough temperatures. The structure and relaxation of diluted
antiferromagnets~\cite{Kleeman97,Natt88} in low external fields are
suitable candidates: there one would presume it to be of relevance
that there are large-scale structures present in the equilibrium
state. In particular in coarsening, it is unclear how the eventual
hull exponent of 4/3 would affect the dynamics. It would be
interesting to see what kind of phenomena can be seen in the structure
on triangular lattices since here $p_c=0.5$ even in the ordinary
site-percolation. One open question or application is the 3D RFIM. The
percolation transition of the minority spins is expected to take place
along a line in the ($H$, $\Delta$) -phase diagram as well, since $p_c
\simeq 0.312$ for site-percolation in the case of the cubic systems
most often studied numerically. Thus in low fields only one of spin
orientations percolates whereas at high fields both do, see a review
of 3D RFIM experiments in~\cite{Belanger00}. The role of this
transition is unclear also when it comes to the ferro- to paramagnet
phase boundary, and the nature of the phase transition.

\section*{Acknowledgements}

This work has been supported by the Academy of Finland Centre of
Excellence Programme. We acknowledge Dr.\ Cristian Moukarzel for
discussions.

%%%%%%%%%%%%%%%%%%%%%%%%%%%%%

%%%%%%%%%%%%%
%FIGURES
%%%%%%%%%%%%%

\begin{figure}[f]
\vspace*{0.5cm} \centerline{\epsfig{file=fig6a.eps,width=7cm}}
\vspace*{1cm} \centerline{\epsfig{file=fig6b.eps,width=7cm}}
\vspace*{1cm} \centerline{\epsfig{file=fig6c.eps,width=7cm}} \vspace*{0.5cm} 
\caption{(a) The susceptibility, calculated as a fluctuation of the 
magnetization, multiplied by the system size $L^2 \chi = L^2 \langle
m^2 - \langle m \rangle^2 \rangle $ versus the external field $H$ for
$\Delta =2.2$.  (b) The scaled susceptibility $\chi/\exp(7.3/\Delta)$
versus the scaled external field $H/\exp(-6.5/\Delta)$ for the system
size $L^2=175^2$, and random field strength values $\Delta =$ 1.9,
2.0, 2.2, 2.6, 3.0, 3.5 and 4.5.  Each point is a disorder-average
over 5000 realizations and the error bars are smaller than the
symbols.  (c) Same as (b) but only the positive external field values
and in lin-log scale. The $\exp(-0.2x)$ line is to guide the eye.}
\label{fig6}
\end{figure}

\begin{figure}[f]
\centerline{\epsfig{file=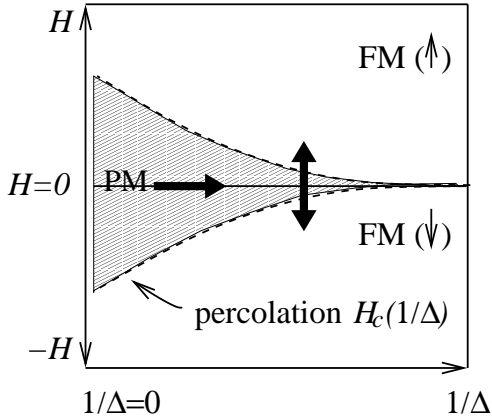,width=8cm,angle=-90}}
\caption{The phase diagram for the 2D RFIM with disorder
strength $\Delta$ and an applied external field $H$.  The $1/\Delta
=0$ axis corresponds to the standard site-percolation, with the
percolation occupation fraction $p_c = 0.593$. The dashed lines define
the percolation thresholds $H_c(1/\Delta)$ for up and down spins,
below and above which the systems are simple ferromagnetic. The thick
arrows denote two directions of which the percolation transition may
be studied: the vertical one fixed $\Delta$ and varying $H$ and the
horizontal one $H=0$ and varying $\Delta$.}
\label{fig7}
\end{figure}

\begin{figure}[f]
\centerline{\epsfig{file=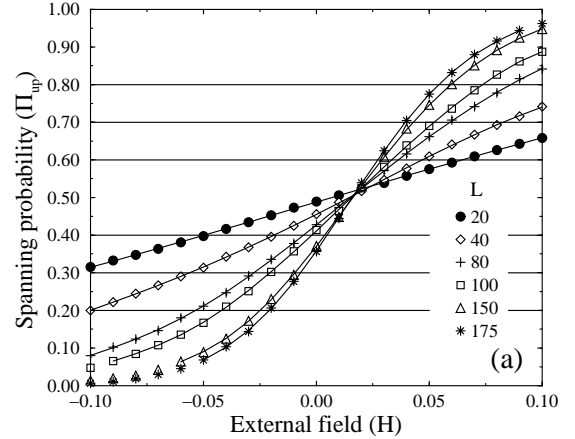,width=6cm,angle=-90}}
\centerline{\epsfig{file=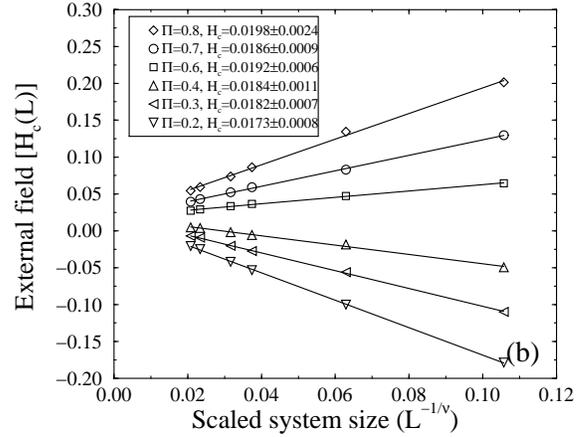,width=6cm,angle=-90}}
\centerline{\epsfig{file=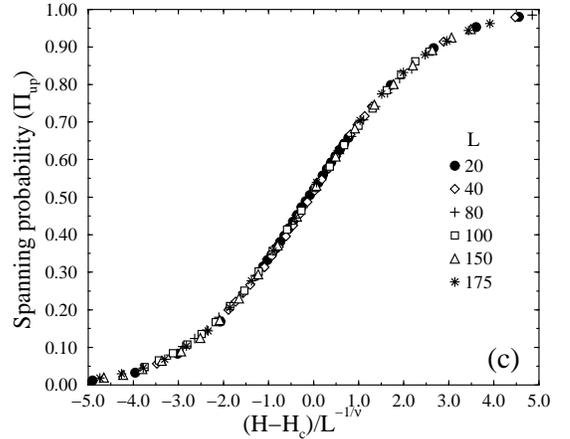,width=6cm,angle=-90}}
\caption{(a) The spanning probabilities of up spins $\Pi_{up}$ as
a function of $H$ for $\Delta =2.6$ with $L^2 \in [20^2\--175^2]$.
The data points are disorder averages over 5000 realizations the error
bars being smaller than the symbols.  The lines are sixth order
polynomial fits. (b) The crossing points $H_c(L)$ of the polynomials
with the horizontal lines leads to the estimate of the critical $H_c$
by finite size scaling using $L^{-1/\nu}$, $\nu =4/3$.  (c) The
data-collapse with the corresponding critical $H_c=0.00186\pm0.0008$.}
\label{fig8}
\end{figure}

\begin{figure}[f]
\centerline{\epsfig{file=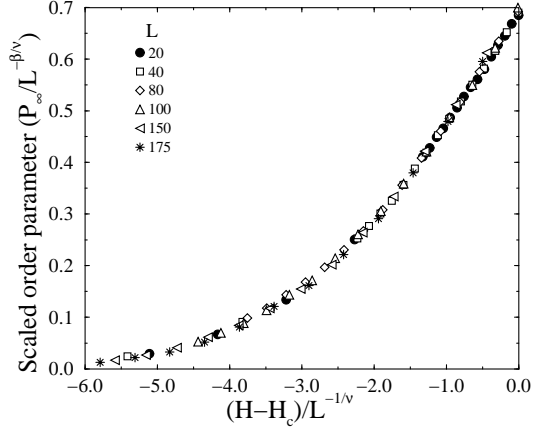,width=6cm,angle=-90}}
\caption{The scaled order parameter, probability of belonging to
the up-spin spanning cluster, $P_\infty/L^{-\beta/\nu}$, $\beta=5/36$,
$\nu =4/3$ versus the scaled external field $(H-H_c)/L^{-1/\nu}$, for
$\Delta =3.0$ with $L^2 \in [20^2\--175^2]$. The data points are
disorder averages over 5000 realizations the error bars being smaller
than the symbols.  The corresponding critical
$H_c(\Delta=3.0)=0.040\pm0.001$.}
\label{fig9}
\end{figure}

\begin{figure}[f]
\centerline{\epsfig{file=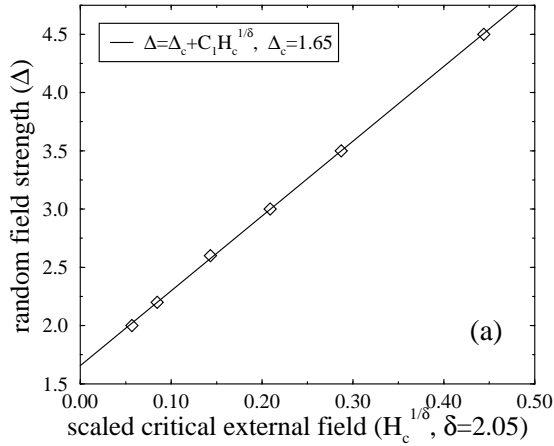,width=6cm,angle=-90}}
\centerline{\epsfig{file=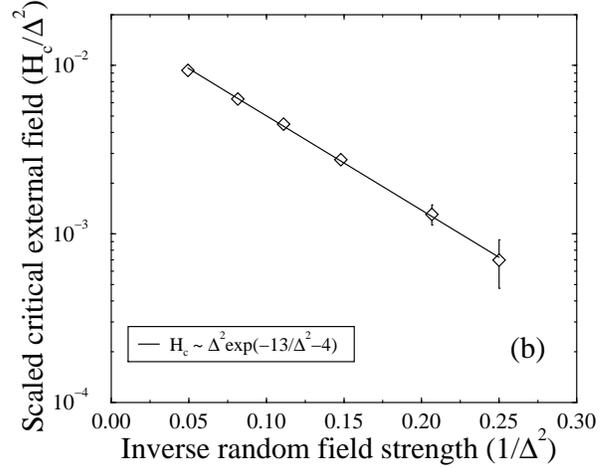,width=7cm,angle=-90}}
\caption{(a) For each $\Delta$ the critical $[H_c(\Delta)]^{1/\delta}$ 
of up spin -spanning, where $\delta=2.05\pm 0.10$. The data follows $H_c 
\sim (\Delta -\Delta_c)^\delta$, where $\Delta_c = 1.65 \pm 0.05$. 
The details are as in Fig.~\ref{fig8}. $H_c (\Delta)$ spans almost two 
decades from $H_c = 0.0028\pm 0.0008$ for $\Delta = 2.0$ to $H_c=0.1891
\pm 0.002$ for $\Delta = 4.5$.  (b) The other scenario is shown, with 
$H_c \sim \Delta^2\exp (-13/\Delta^2 -4)$ (see text).}
\label{fig10}
\end{figure}

\begin{figure}[f]
\centerline{\epsfig{file=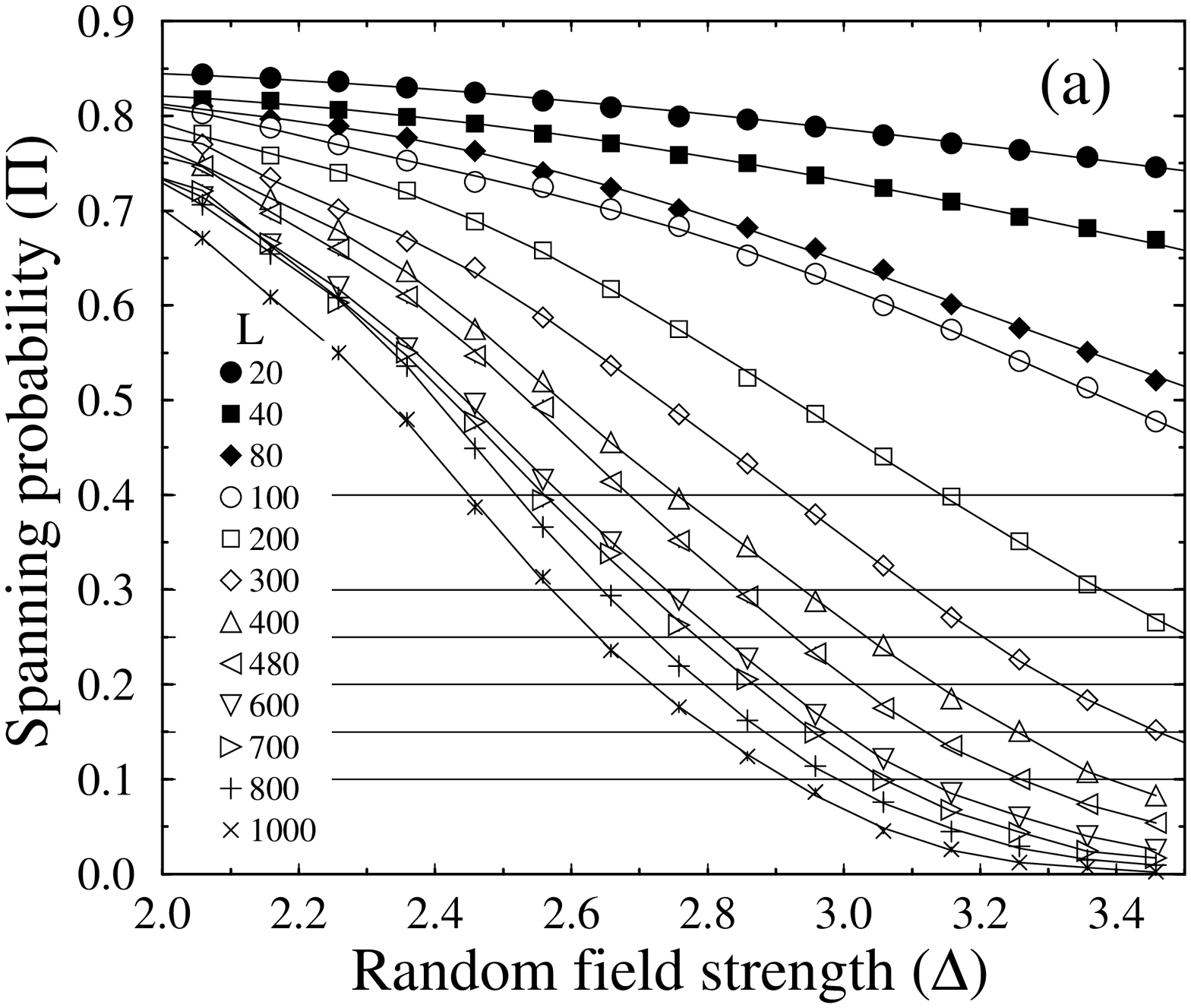,width=6cm,angle=-90}}
\centerline{\epsfig{file=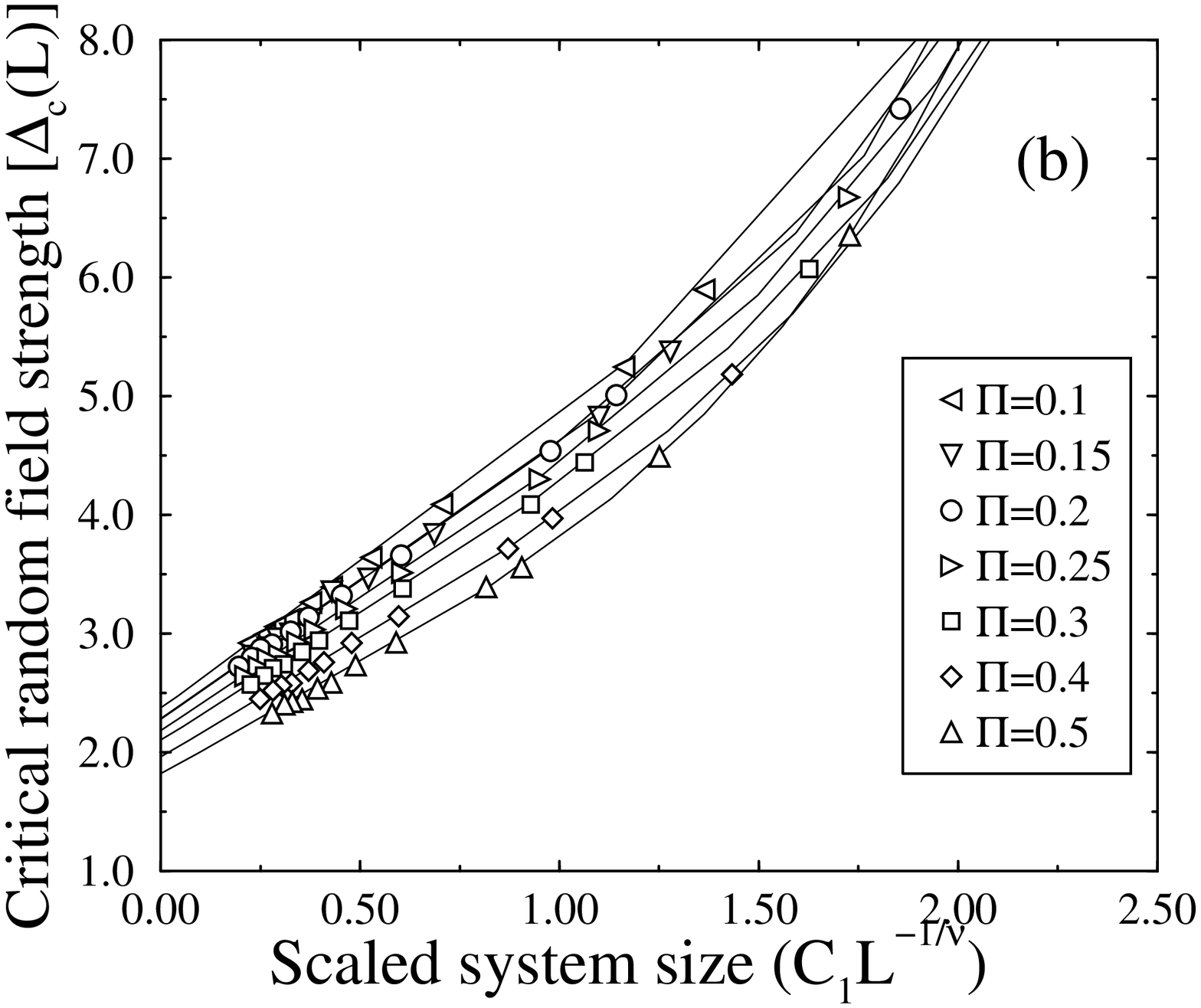,width=6cm,angle=-90}}
\centerline{\epsfig{file=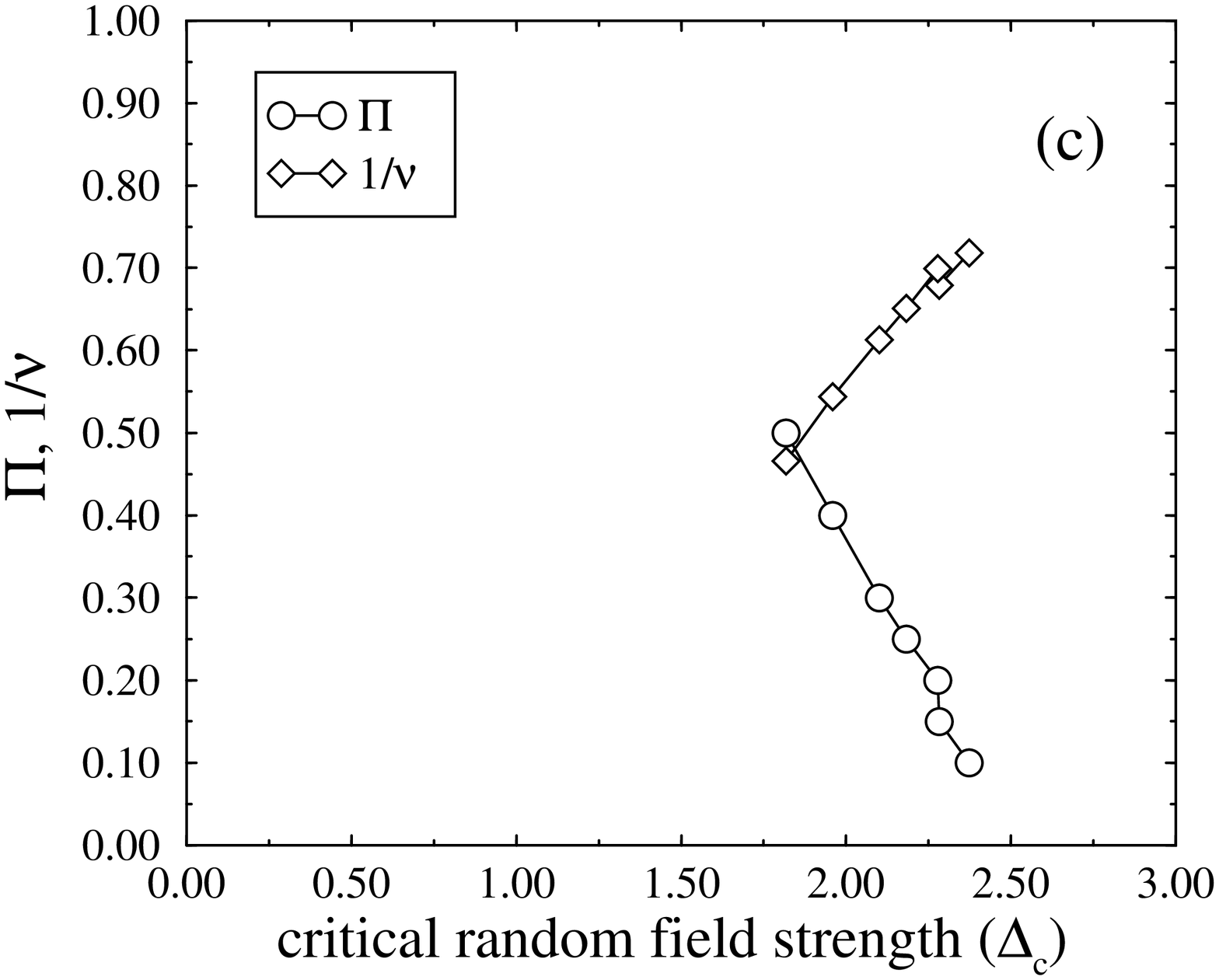,width=6cm,angle=-90}}
\centerline{\epsfig{file=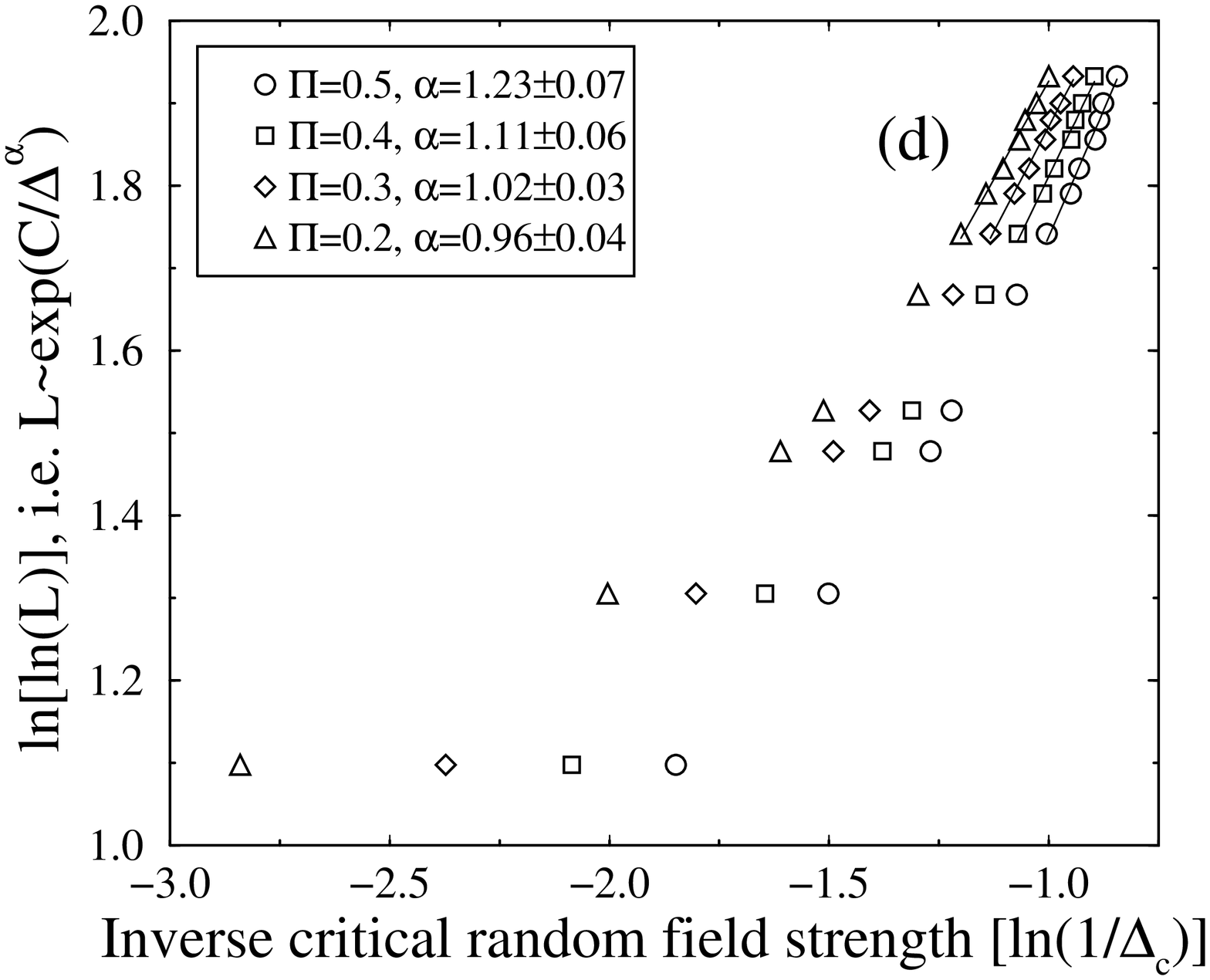,width=6cm,angle=-90}}
\caption{(a) The spanning probabilities of up or down spins $\Pi$ as
a function of $\Delta$ for $H=0$ with $L^2 \in [20^2\--1000^2]$.  The
data points are disorder averages over 5000 realizations the error
bars being smaller than the symbols.  The lines are tenth order
polynomial fits. (b) The crossing points $\Delta_c(L)$ of the
polynomials with the horizontal lines of $\Pi=$0.1, 0.15, 0.2, 0.25,
0.3, 0.4, and 0.5 versus the scaled system size. The estimates of the
critical $\Delta_c$ are searched by finite size scaling using
$\Delta_c(L) = \Delta_c(1+C_1L^{-1/\nu})(1+C_2L^{-1/\nu_2})$ and the
data is plotted as $\Delta_c(L)$ versus $C_1L^{-1/\nu}$.  (c) The
critical random field values $\Delta_c$ with respect the $\Pi$ values
they are estimated from. The corresponding correlation length
exponents $1/\nu$, which are used in (b), are shown, too.  (d) Another
scenario: The critical random field values $\Delta_c$ with respect to
system size. The lines are least-squares fits of form $L \sim
\exp(C/\Delta^\alpha)$, where $C$ is a free parameter, for different
$\Pi$ values from (a).}
\label{fig11}
\end{figure}

\begin{figure}[f]
\centerline{\epsfig{file=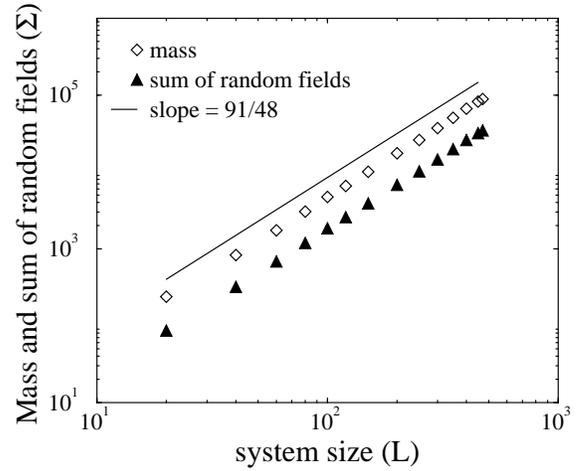,width=7cm,angle=-90}}
\caption{The average mass of spanning clusters for bimodal field randomness 
$\Delta= 25/13$ up to system size $L=470$. The plot shows also the sum
of the random fields of the sites belonging to the same clusters.  The
2d percolation fractal dimension $D =$91/48 is indicated with a line.}
\label{fig12}
\end{figure}

\begin{figure}[f]
\centerline{\epsfig{file=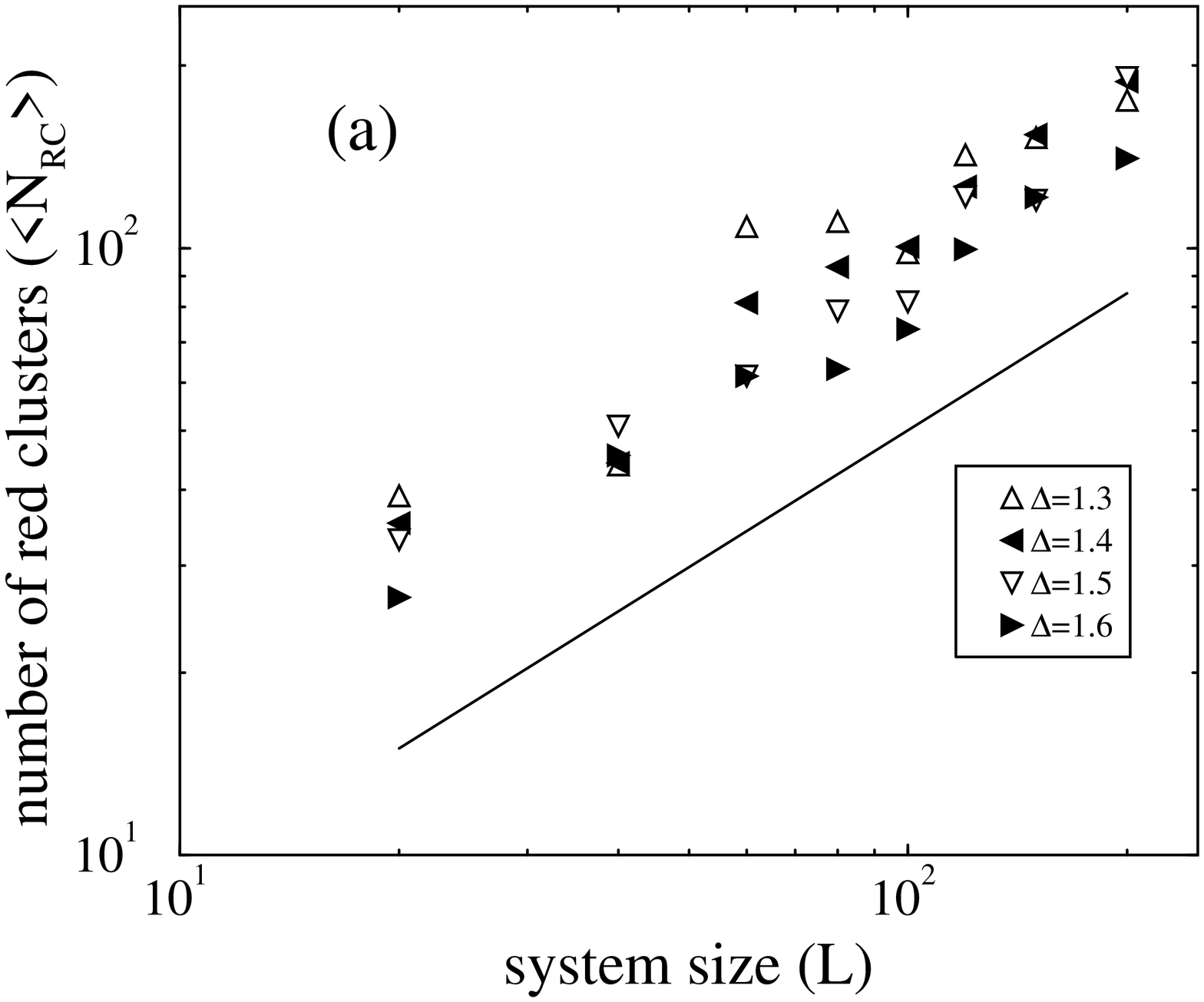,width=6cm,angle=-90}}
\centerline{\epsfig{file=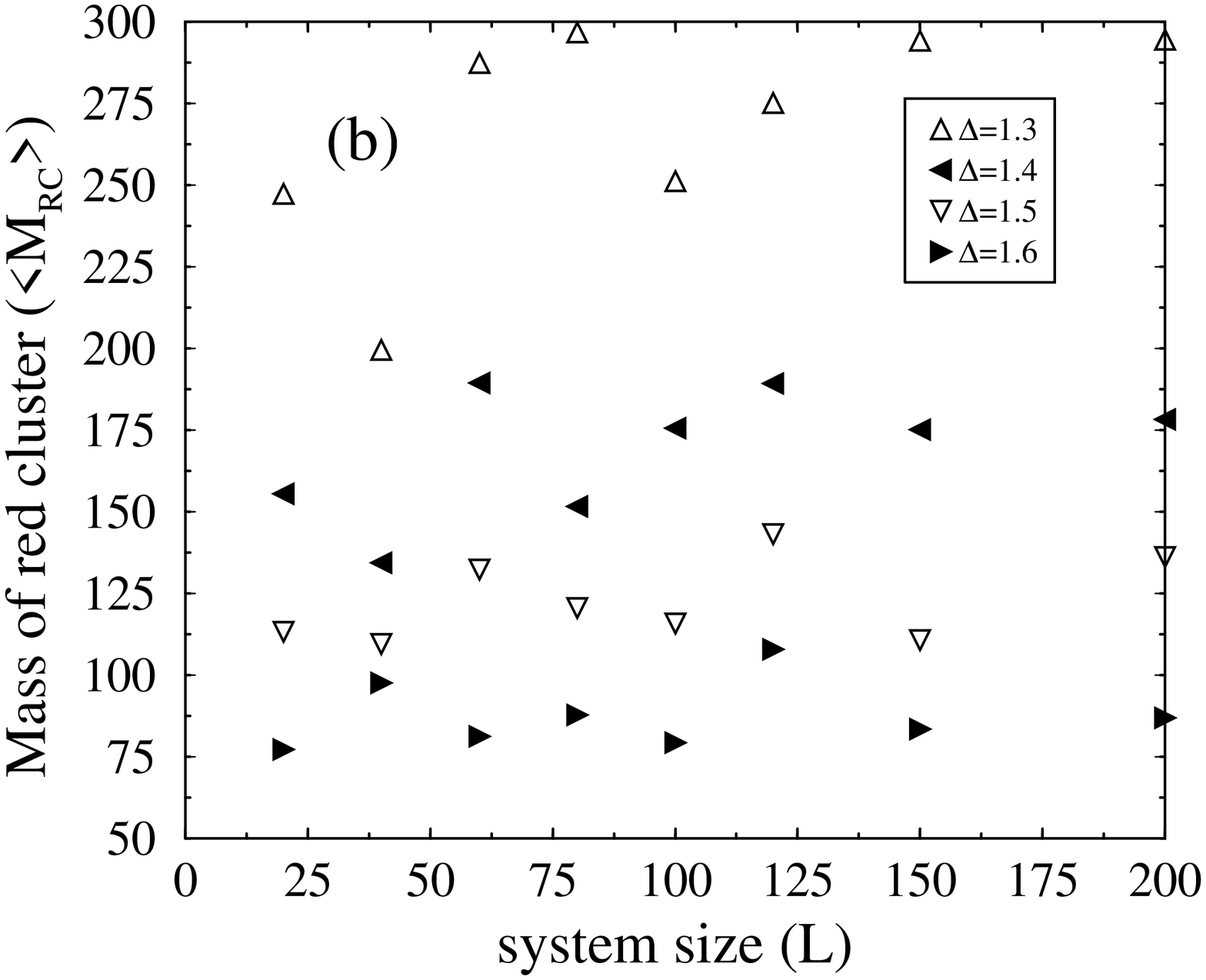,width=6cm,angle=-90}}
\caption{(a) The average number of red clusters, $\langle N_{RC} \rangle$, 
as a function of the system size, $L^2= 20^2 \-- 200^2$, for $\Delta=$
1.3, 1.4, 1.5, and 1.6 with $H=0$. Here the number of realizations is
$N=200$.  The smaller the $\Delta$ the larger the amplitude of
$\langle N_{RC}\rangle$, since the sizes of red clusters become
larger.  The line $L^{3/4}$ is a guide to the eye. (b) The masses of
the red clusters $\langle M_{RC} \rangle$ with respect to the system
size $L$. $\langle M_{RC} \rangle$ does not depend on the system size
as seen in the figure.}
\label{fig13}
\end{figure}

\end{multicols}
\widetext
\end{document}